\documentclass[twocolumn,floatfix,amsmath,amssymb,aps]{revtex4}
\usepackage{epsfig}

\begin{document}

\title{Field theory of bicritical and tetracritical points. IV. Critical dynamics
including reversible terms.}

\author{R. Folk}\email{folk@tphys.uni-linz.ac.at}
\affiliation{Institute for Theoretical Physics, Johannes Kepler
University Linz, Altenbergerstrasse 69, A-4040, Linz, Austria}
 \author{Yu. Holovatch}\email[]{hol@icmp.lviv.ua}
 \affiliation{Institute for Condensed Matter Physics, National
Academy of Sciences of Ukraine, 1~Svientsitskii Str., UA--79011
Lviv, Ukraine}\affiliation{Institute for Theoretical Physics, Johannes Kepler
University Linz, Altenbergerstrasse 69, A-4040, Linz, Austria}
  \author{G. Moser}\email[]{guenter.moser@sbg.ac.at}
\affiliation{Department for Material Research and Physics, Paris Lodron University
Salzburg, Hellbrunnerstrasse 34, A-5020 Salzburg, Austria}
\date{\today}
\begin{abstract}
This article concludes a series of papers (R. Folk, Yu. Holovatch,
and G. Moser, Phys. Rev. E {\bf 78}, 041124 (2008); {\bf 78}, 041125
(2008); {\bf 79}, 031109 (2009)) where the tools of the field
theoretical renormalization group were employed to explain and
quantitatively describe different types of static and dynamic
behavior in the vicinity of multicritical points. Here, we give the
complete two loop calculation and analysis of the dynamic
renormalization-group flow equations at the multicritical
point in anisotropic antiferromagnets in an external magnetic field.
We find that the time scales of the order parameters  characterizing the parallel and
perpendicular ordering with respect to the external field scale in
the same way. This holds independent whether the
Heisenberg fixed point or the biconical fixed point in statics is
the stable one. The non-asymptotic analysis of the dynamic flow
equations shows that due to cancelation effects the critical
behavior is described - in distances from the critical point
accessible to experiments - by the critical behavior qualitatively
found in one loop order. Although one may conclude from the
effective dynamic exponents (taking almost their one loop values)
that weak scaling for the order parameter components is valid,
the flow of the time scale ratios is quite different  and they do
not reach their asymptotic values.
\end{abstract}

\maketitle

\section{Introduction}

Three component antiferromagnets in three spatial dimensions in an
external magnetic field in $z$ direction contain in their phase
diagram two second order transition lines: (i) between the
paramagnetic and the spin flop phase and (ii) between the
antiferromagnetic  and paramagnetic phase. The point where these two
lines meet is a multicritical point which turned out to be either
bicritical or tetracritical. Within the renormalization group (RG) theory the stability and attraction region of the static fixed point (FP) of the RG transformation determines, which kind of multicritical behavior is realized. For the bicritical point it is the Heisenberg FP, for the tetracritical point it is the biconical one. The stabilty of a FP depends on the system's global features  as the space and order parameter (OP) dimensions $d$ and $n$. In $d=3$, the case considered here, the biconical FP is stable apart from a  restricted attraction region of the Heisenberg FP .  The static phase transition on each of
the phase transition lines belongs for (i) to an isotropic
Heisenberg model with $n_\perp=2$ and for (ii) to Heisenberg
model with $n_\|=1$ \cite{konefi76,partI}.

Concerning the dynamical universality classes the transition
(i) belongs to the class described by model F and (ii) belongs to
the model C class (for the notation see \cite{review}). At the
multicritical point the critical behavior is described by a new
universality class both in statics and dynamics. The interesting
feature of these systems is that all the different OPs
 characterizing the ordered phase are physically accessible.
This is most important for the dynamical behavior since the only
other  example belonging to model F is the superfluid transition in
$^4$He where the OP (the complex macroscopic wave function of the
condensate \cite{josephson66}) is experimentally not accessible \cite{hh69}.
Here the OPs are the components of the staggered magnetization.
Their correlations (static and dynamical) can be measured by neutron
scattering.

A complete description of the critical dynamics near the multicritical point mentioned
above has to take into account the slow dynamical densities which
are the OPs and the conserved densities present in the system. Due
to the  external magnetic field the only conserved density which has
to be taken into account is the magnetization in direction of the
external field. A derivation of the dynamical equations follows
along the usual steps calculating the reversible terms from the
non-zero Poisson brackets, introducing irreversible terms present
also in the hydrodynamic limit, dropping irrelevant terms and taking
into account terms arising in the renormalization procedure (see
e.g. the review \cite{review}). Such a dynamical model has already
been considered in \cite{dohmjanssen77,dohmmulti83,dohmKFA}  by
RG theory and it was argued that due to
nonanalytic terms in $\varepsilon=4-d$ a FP in two
loop order qualitative different from the one loop FP is found. The
result of the one loop calculations is that the time scales of the parallel and
perpendicular components of the staggered magnetization scale
differently whereas calculated in two loop order they scale similar
although the FP value of the timescale ratio of the two components
cannot be found by $\varepsilon$ expansion and might be very small
in $d=3$ namely of ${\cal O}(10^{-86})$. It was argued that the
terms leading to the singular behavior in $\varepsilon$ do not
contribute to the FP value of the mode coupling. The calculations of the
RG-functions in
\cite{dohmjanssen77} where not complete in two loop order (they took
 into account only the terms which lead to the nonanalytic behavior
in $\varepsilon$). At that time also the Heisenberg FP (named
$\cal{H}$) was considered to be the stable static one, whereas it
turned out in two loop order (resummed) that it is the biconical FP
(named $\cal{B}$) \cite{partI}.

\begin{table}[h]
\centering \tabcolsep=3mm
\begin{tabular}{c|c|c|c|c}
 \hline \hline
FP values in ... & $w^\star_\|$& $w^{\prime\star}_\perp$ & $v^{\prime\star}$ &  $f^\star_\perp$ \\ \hline \hline
${\cal B_C}$  \cite{partIII} & $0.76$ & $\gg 1$ &$\sim 0$ & -\\  \hline
${\cal B}$ 1-loop & $0$ & $1.555$ & $0$ & $1.086$\\ \hline
${\cal B}$ 2-loop & $0$ & $0$ & $\sim 0$& $1.131$ \\ \hline
model C \cite{C} & $0.49$ & - & - & - \\ \hline
model F \cite{F} & - & 0 & - & $0.834$ \\ \hline
\end{tabular}
\caption{Dynamical FP values (zeros of the corresponding dynamical
$\beta$-functions)  at $d=3$ of different models for the time scale ratios
$w^\star_\|, \, w^{\prime\star}_\perp, \, v^{\prime\star}$ and the
mode coupling constant $f^\star_\perp$. The 2nd and 3rd lines quote
results of this paper found in the biconical static FP for the
tetracritical behavior of the dynamical model that takes into
account reversible terms. They are compared with the two loop
results found in the model C multicritical point \cite{partIII} as
well as in the critical points of model C for the one component OP
\cite{C} and of model F for the two component order parameter
\cite{F}. \label{tab1}}
\end{table}

A summary of the  results obtained so far for the FPs characterizing
dynamical behavior  is given in table \ref{tab1}. Neglecting the
reversible terms one is left with a purely relaxational dynamics.
Then the asymptotic dynamical critical behavior is characterized by
the FP values of the independent time scale ratios of the system.
These are the following time scale ratios: (1) the ratio $w_\|$
between the relaxation rate of the staggered magnetization parallel
to the external field and the diffusive transport coefficient of the
magnetization parallel to the external field; (2) the
 ratio
$w^\prime_\perp$ between the real part of the relaxation rate of the
staggered magnetization perpendicular to the external field and the
diffusive transport coefficient of the magnetization parallel to the
external field. In addition we introduce the ratio $v^\prime$
between the two components of the real relaxation rates of the two
OPs in order to compare their dynamic scaling behavior. A non-zero
finite value of the time scale ratio means that the two involved
densities scale with the same exponent. If all time scale ratios are
non-zero and finite, one speaks of strong dynamic scaling, otherwise of weak
dynamic scaling. Especially of interest is the behavior of the scaling of
the two components of the  OP indicated by the FP value of
$v^\prime$. In the third line of table \ref{tab1} the two loop order
result shows weak dynamic scaling between the OPs and the conserved
density but strong scaling between the OP components. However since
the FP value time scale ratio $v^\prime$ is almost zero, the
critical behavior is dominated by non-asymptotic effects. For
comparison the FP values for the case of model C for the one
component OP \cite{C} and for model F for the two component order
parameter \cite{F} are included. They are the limiting cases when
the two OPs characterizing the multicritical behavior would decouple
in statics and dynamics.

In the first line of table \ref{tab1} the results for the
multicritical dynamical FP  $\cal{B}_C$ values taking into
account the static coupling of the OP to the conserved density
are displayed (see \cite{partIII}). All time scale ratios are
non-zero and finite but since $w^{\prime\star}_\perp$ is very large
($v^{\prime\star}$ almost zero) the observable behavior in the
vicinity of the multicritical point is predicted to be dominated by
non-asymptotic effects and strong scaling is not observable
\cite{partIII}.  In the second line the results of a one loop RG
calculation with reversible terms for the biconical FP are given.
The FP value of the mode coupling parameter $f_\perp$ is finite but
since $w^\star_\|=0$ the critical dynamics is characterized by weak
dynamic scaling and the two components of the OP scale different. A
similar result for the Heisenberg FP was found in
\cite{dohmjanssen77}.  In the third line the results found in this
paper are shown, indicating weak scaling between the conserved
density and the components of the OP, but strong scaling between the
parallel and perpendicular components of the OP. Since the FP value
of the time scale ratio between the component $v^{\prime\star}$ is
almost zero but definitively different from zero it is expected that
non-asymptotic behavior is dominant.

This article concludes a series of papers
\cite{partI,partII,partIII} (henceforth cited as papers I, II, and
III) where the tools of the field theoretical RG were employed to
explain and quantitatively describe different types of static and
dynamic behavior in the vicinity of multicritical points. A short account of the results presented here was given in \cite{letter}. The
statics and dynamics were treated in Refs. \cite{partI} and
\cite{partII,partIII}, respectively. First, purely relaxational
dynamics was considered (paper II) and later, in paper III, these
results served as a basis to consider how an account of
magnetization conservation modifies dynamical behavior. The goal of
the current study is more ambitious: we will analyze a complete set of
dynamical equations of motion taking into account reversible terms \cite{huber74,huberrag76}
and give a comprehensive description of dynamical behavior
in the vicinity of multicritical points in two loop order. The paper
is organized as follows: In section \ref{equation} the dynamic model
is defined followed by a the definitions of the dynamical functions
considered in section \ref{vertex}. The renormalization and
corresponding RG-functions are presented in section \ref{renorm} and
\ref{rngfunctions} respectively. The two loop results of our
calculations for these dynamic RG-functions are given in section
\ref{twoloop}. The one loop approximation for the dynamic is
discussed in section \ref{onelooporder}. In the next section
\ref{limitzeta} we consider the asymptotic properties of the two
loop RG-functions leading to the general asymptotic results in
section \ref{generalbehavior}. We then present the  results expected
in the asymptotic subspace, section \ref{asymptotics}. The
non-asymptotic behavior,  obtained by looking at the region further away from the multicritical point, is
shown in section \ref{pseudoasymptotics}, a conclusion
\ref{conclusion} ends the paper. In Appendices  calculational
details for some intermediate steps of the RG calculation are
presented.

\section{Model equations of the antiferromagnet in an external field} \label{equation}

The non-conserved OP $\vec{\phi}_0$ of an isotropic
antiferromagnet is given by the three dimensional vector
\begin{equation}\label{opvec}
\vec{\phi}_0=\left(\begin{array}{c} \phi_0^x \\
\phi_0^y \\ \phi_0^z \end{array}\right)
\end{equation}
of the staggered magnetization, which is the difference of two
sublattice magnetizations. An external magnetic field applied to the
ferromagnet induces an anisotropy to the system. The OP splits into
two OPs, $\vec{\phi}_{\perp 0}$ perpendicular to the field, and
$\vec{\phi}_{\| 0}$ parallel to the external field. Assuming the
$z$-axis in direction of the external magnetic field, the two OPs are
\begin{equation}\label{opsep}
\vec{\phi}_{\perp 0}=\left(\begin{array}{c} \phi_0^x \\
\phi_0^y \end{array}\right)  \ , \qquad
\phi_{\| 0}=\phi_0^z
\end{equation}
In addition the $z$-component of the magnetization $m_0$ has to
be taken into account for the dynamics and therefore has to be
included in statics although there it could be integrated out and
does not change the asymptotic static critical behavior. Thus the
static critical behavior of the system is described by the
functional
\begin{eqnarray}\label{1}
{\cal H}\!=\!\int\!
d^dx\Bigg\{\frac{1}{2}\mathring{r}_\perp\vec{\phi}_{\perp 0}
\cdot\vec{\phi}_{\perp
0}+\frac{1}{2}\sum_{i=1}^{d}\nabla_i\vec{\phi}_{\perp 0}\cdot
\nabla_i\vec{\phi}_{\perp 0}
\nonumber \\
+\frac{1}{2}\mathring{r}_\|{\phi}_{\| 0} {\phi}_{\|
0}+\frac{1}{2}\sum_{i=1}^{d}\nabla_i\phi_{\| 0}
\nabla_i\phi_{\| 0}
+\frac{\mathring{u}_\perp}{4!}\Big(\vec{\phi}_{\perp
0}\cdot\vec{\phi}_{\perp 0}\Big)^2 \nonumber \\
+\frac{\mathring{u}_\|}{4!}\Big(\phi_{\| 0}\phi_{\|
0}\Big)^2
+\frac{2\mathring{u}_\times}{4!}\Big(\vec{\phi}_{\perp
0}\cdot\vec{\phi}_{\perp 0}\Big) \Big(\phi_{\|
0}\phi_{\| 0}\Big) \Bigg\}  \\
+\frac{1}{2}m_0^2+\frac{1}{2}\mathring{\gamma_\perp}m_0\vec{\phi}_{\perp
0}\cdot\vec{\phi}_{\perp 0}
+\frac{1}{2}\mathring{\gamma_\|}m_0\phi_{\|
0}\phi_{\| 0} -\mathring{h}m_0\Bigg\}  \ , \nonumber
\end{eqnarray}
with familiar notations for bare couplings $\{\mathring{u},\mathring{\gamma}\}$,
masses $\{\mathring{r}\}$ and field $\mathring{h}$ \cite{partI,partII}.
The critical dynamics of relaxing OPs coupled to a diffusing
secondary density is governed by the following equations of motion
\cite{dohmjanssen77}:
\begin{eqnarray}
\label{dphiperp}
\frac{\partial \phi_{\perp 0}^\alpha}{\partial
t}&=&-\mathring{\Gamma}^\prime_\perp \frac{\delta {\mathcal
H}}{\delta \phi_{\perp 0}^\alpha}
+\mathring{\Gamma}^{\prime\prime}_\perp \epsilon^{\alpha\beta z}
\frac{\delta {\mathcal H}}{\delta \phi_{\perp 0}^\beta} \nonumber \\
&+&\mathring{g}\ \epsilon^{\alpha\beta z}
\phi_{\perp 0}^\beta\frac{\delta {\mathcal H}} {\delta m_0}+\theta_{\phi_\perp}^\alpha
\ ,
\end{eqnarray}
\begin{eqnarray}
\label{dphipar}
\frac{\partial \phi_{\|0}}{\partial t}&=&-\mathring{\Gamma}_\|
\frac{\delta {\mathcal H}}{\delta \phi_{\|0}}+\theta_{\phi_\|}
\, , \\
\label{dmpar}
\frac{\partial m_0}{\partial t}&=&\mathring{\lambda}\nabla^2
\frac{\delta {\mathcal H}}{\delta m_0}+
\mathring{g}\ \epsilon^{\alpha\beta z}\phi_{\perp 0}^\alpha
\frac{\delta {\mathcal H}}{\delta\phi_{\perp 0}^\beta}
 +\theta_m \, ,
\end{eqnarray}
with the Levi-Civita symbol $\epsilon^{\alpha \beta z}$.
Here $\alpha,\beta=x,y$ and the sum over repeated indices is implied.

The dynamical equations describe the dynamics of an antiferromagnet
with the usual Lamor precession terms for the alternating
magnetization and relaxational terms. Due to the static
 coupling to the conserved magnetization additional Lamor terms arise together with a diffusive
term for the magnetization. Renormalization considerations lead on
one hand to a neglection of several Lamor terms and on the other
hand create an additional reversible term (the second term on the
right hand side of Eq. \ref{dphipar})) not present in the usual
dynamics of antiferromagnets \cite{antiferro}.

Combining the kinetic coefficients of the OP to a complex quantity,
$\mathring{\Gamma}_\perp=\mathring{\Gamma}_\perp^\prime
+\mbox{i}\mathring{\Gamma}_\perp^{\prime\prime}$, the imaginary part
constitutes a precession term created by the renormalization
procedure even if it is absent in the background. The kinetic
coefficient $\mathring{\lambda}$ and the mode coupling
$\mathring{g}$ are real. The stochastic forces
$\vec{\theta}_{\phi_\perp}$, $\vec{\theta}_{\phi_\|}$ and $\theta_m$
fulfill Einstein relations
\begin{eqnarray} \label{thetaperp}
\langle\theta_{\phi_\perp}^\alpha(\vec{x},t)\ \theta_{\phi_\perp}^\beta
(\vec{x}^\prime,t^\prime)\rangle \!\!\!&=&\!\!\!
2\mathring{\Gamma}^\prime_\perp\delta(\vec{x}-\vec{x}^\prime)\delta(t-t^\prime)\delta^{\alpha\beta}
\ , \\
\label{thetapara}
\langle\theta_{\phi_\|}(\vec{x},t)\ \theta_{\phi_\|}(\vec{x}^\prime,t^\prime)
\rangle \!\!\!&=&\!\!\! 2\mathring{\Gamma}_\|\delta(\vec{x}-\vec{x}^\prime)\delta(t-t^\prime)
\ , \\
\label{thetam}
\langle\theta_m(\vec{x},t)\ \theta_m(\vec{x}^\prime,t^\prime)
\rangle \!\!\!&=&\!\!\! -2\mathring{\lambda}\nabla^2\delta(\vec{x}-\vec{x}^\prime)\delta(t-t^\prime)
\ .
\end{eqnarray}

In view of dynamical calculations it is more convenient to deal with
a scalar complex order parameter
$\psi_0=\psi_0^\prime+\mbox{i}\psi_0^{\prime\prime}$ instead of the
real two-dimensional OP $\vec{\phi}_{\perp 0}$ in
(\ref{opsep}). Thus we may introduce
\begin{equation}\label{psidef}
\psi_0=\phi_0^x -\mbox{i}\phi_0^y \ , \qquad  \psi_0^+=\phi_0^x +\mbox{i}\phi_0^y
\end{equation}
as OP of the perpendicular components. The superscript $^+$ denotes
complex conjugated quantities also in the following equations. In
addition to the two OPs the $z$-component of the magnetization, which
is the sum of the two sublattice magnetizations, has to be
considered as conserved secondary density $m_0$.

Expressed in terms of the above densities the dynamic equations take the form
\begin{eqnarray}
\label{dphidt}
\frac{\partial \psi_0}{\partial t}\!\!&=&\!\!-2\mathring{\Gamma}_\perp
\frac{\delta H}{\delta \psi_0^+}+\mbox{i}\psi_0 \mathring{g}
\frac{\delta H}{\delta m_0}+\theta_\psi \, ,\\
\label{dphiccdt}
\frac{\partial \psi_0^+}{\partial t}\!\!&=&\!\!-2\mathring{\Gamma}_\perp^+
\frac{\delta H}{\delta \psi_0}-\mbox{i}\psi_0^+\mathring{g}
\frac{\delta H}{\delta m_0}+\theta_\psi^+ \, ,\\
\label{dphipadt}
\frac{\partial \phi_{\|0}}{\partial t}\!\!&=&\!\!-\mathring{\Gamma}_\|
\frac{\delta H}{\delta \phi_{\|0}}+\theta_{\phi_\|} \, ,\\
\label{dmdt} \frac{\partial m_0}{\partial t} \!\!&=&\!\!
\mathring{\lambda}\nabla^2\frac{\delta H}{\delta m_0}-2\mathring{g}
\Im[\psi_0^+\nabla^2\psi_0]+\theta_m \, .
\end{eqnarray}
Due to the fact that the stochastic forces
$\theta_{\phi_\perp}^\alpha$ in (\ref{dphiperp}) are
$\delta$-correlated and fulfil the Einstein relations, similar
properties hold also for the stochastic forces $\theta_{\psi}$:
\begin{eqnarray}\label{einstein}
\langle\theta_\psi(x,t)\ \theta_\psi^+(x',t')\rangle =
4\mathring{\Gamma}_\perp^\prime\ \delta(x-x')\delta(t-t') \, .
\end{eqnarray}
The critical behavior of the
thermodynamic derivatives follows from the extended static functional (the functional (\ref{1}) written in the variables intoduced in (\ref{psidef}))
\begin{equation}\label{statfunc}
{\cal H}={\cal H}^{(0)}+{\cal H}^{(int)}
\end{equation}
with a Gaussian part
\begin{eqnarray}\label{h0}
{\cal H}^{(0)}\!=\!\int\! d^dx\Bigg\{\frac{1}{2}\mathring{\tilde{r}}_\perp\psi_0^+
\psi_0+\frac{1}{2}(\nabla\psi_0^+)(\nabla\psi_0)  \nonumber \\
+\frac{1}{2}\mathring{\tilde{r}}_\|\phi_{\| 0}^2
+\frac{1}{2}(\nabla\phi_{\| 0})^2+\frac{1}{2}m_0^2
-\mathring{h}m_0\Bigg\} \ ,
\end{eqnarray}
and an interaction part
\begin{eqnarray}\label{hint}
{\cal H}^{(int)}\!=\!\int\! d^dx\Bigg\{
\frac{\mathring{\tilde{u}}_\perp}{4!}(\psi_0^+\psi_0)^2
+\frac{\mathring{\tilde{u}}_\|}{4!}\phi_{\| 0}^4
+\frac{2\mathring{\tilde{u}}_\times}{4!}\psi_0^+\psi_0\phi_{\|0}^2 \nonumber \\
+\frac{1}{2}\mathring{\gamma}_\perp m_0\psi_0^+\psi_0
+\frac{1}{2}\mathring{\gamma}_\|m_0\phi_{\| 0}^2  \Bigg\} \ .  \nonumber \\
\end{eqnarray}
The above static functional may be reduced to the
Ginzburg-Landau-Wilson (GLW) functional with complex OP by
considering an appropriate Boltzmann distribution and
integrating out the secondary density. One obtains
\begin{eqnarray}\label{hbiglw}
{\cal H}_{GLW}\!=\!\int\! d^dx\Bigg\{\frac{1}{2}\mathring{r}_\perp\psi_0^+
\psi_0+\frac{1}{2}(\nabla\psi_0^+)(\nabla\psi_0)  \nonumber \\
+\frac{1}{2}\mathring{r}_\|\phi_{\| 0}^2
+\frac{1}{2}(\nabla\phi_{\| 0})^2 \nonumber \\
+\frac{\mathring{u}_\perp}{4!}(\psi_0^+\psi_0)^2
+\frac{\mathring{u}_\|}{4!}\phi_{\| 0}^4
+\frac{2\mathring{u}_\times}{4!}\psi_0^+\psi_0\phi_{\|0}^2 \Bigg\} \ .
\end{eqnarray}
The parameters $\{\mathring{r}\}\equiv\mathring{r}_\perp,\mathring{r}_\|$ and
$\{\mathring{u}\}\equiv\mathring{u}_\perp,\mathring{u}_\|,\mathring{u}_\times$
in (\ref{hbiglw}) are related
to the corresponding parameters of the extended static functional (\ref{statfunc}) by
\begin{eqnarray}
\label{rupperprel}
\mathring{r}_\perp=\mathring{\tilde{r}}_\perp+\mathring{\gamma}_\perp\mathring{h}
\ , \qquad  &&
\mathring{u}_\perp=\mathring{\tilde{u}}_\perp-3\mathring{\gamma}_\perp^2  \ ,
\\
\label{rupararel}
\mathring{r}_\|=\mathring{\tilde{r}}_\|+\mathring{\gamma}_\|\mathring{h}
\ , \qquad   &&
\mathring{u}_\|=\mathring{\tilde{u}}_\|-3\mathring{\gamma}_\|^2  \ ,
\\
\label{utimesrel}
&&\mathring{u}_\times=\mathring{\tilde{u}}_\times-3\mathring{\gamma}_\perp
\mathring{\gamma}_\|
\ .
\end{eqnarray}
The property that the static critical behavior does not depend on
the secondary densities, which can be integrated out in
(\ref{statfunc}), leads to relations between the correlation
functions of the secondary densities and the OP correlation
functions. These relations and their derivations have been
extensively discussed in paper III with real OP functions
$\vec{\phi}_{\perp 0}$ and $\phi_{\| 0}$. Because the derivation of
the relations is independent of the type of OP (real or complex),
all of the relations remain valid and can be taken over from paper
III. Therefore we will not repeat them here.

\section{Dynamic correlation and vertex functions} \label{vertex}

The Fourier transformed dynamic correlation functions of the two OPs are usually
introduced as
\begin{equation}\label{cpsipsixt}
\mathring{C}_{\psi\psi^{+}}(\{\xi\},k,\omega)=\!\! \int\!\! d^dx\!\! \int\!\! dt
e^{-ikx+i\omega t} \langle\psi_0(x,t)\psi_0^+(0,0)\rangle_c
\end{equation}
\begin{equation}\label{cphiphixt}
\mathring{C}_{\phi_{\|}\phi_{\|}}(\{\xi\},k,\omega)=\!\! \int\!\! d^dx\!\! \int\!\! dt
e^{-ikx+i\omega t} \langle\phi_{\|0}(x,t)\phi_{\|0}(0,0)\rangle_c
\end{equation}
All functions depend on the two correlation lengths $\xi_\perp$ and
$\xi_\|$, which is indicated by $\{\xi\}$ in a short notation.
$\langle A B\rangle_c=\langle A B\rangle-\langle A \rangle\langle  B\rangle$ denotes the cumulant.
The averages are calculated with a propability density including a dynamic functional, which can be constituted from
the dynamic equations (\ref{dphidt}) - (\ref{dmdt}). In the considered approach of \cite{bauja76} for every density
auxiliary densities are introduced accordingly. They are denoted as $\tilde{\psi}_0^+$, $\tilde{\psi}_0$,
$\tilde{\phi}_{\|0}$ and $\tilde{m}_0$.
The dynamic correlation functions of the order parameters are connected to dynamic vertex functions via
\begin{equation}\label{cpsipsivert}
\mathring{C}_{\psi\psi^{+}}(\{\xi\},k,\omega)=
-\frac{\mathring{\Gamma}_{\tilde{\psi}\tilde{\psi}^+}(\{\xi\},k,\omega)}
{\left\vert\mathring{\Gamma}_{\psi\tilde{\psi}^+}(\{\xi\},k,\omega)\right\vert^2}
\, ,
\end{equation}
\begin{equation}\label{cphiphivert}
\mathring{C}_{\phi_{\|}\phi_{\|}}(\{\xi\},k,\omega)=
-\frac{\mathring{\Gamma}_{\tilde{\phi}_{\|}\tilde{\phi}_{\|}}(\{\xi\},k,\omega)}
{\left\vert\mathring{\Gamma}_{\phi_{\|}\tilde{\phi}_{\|}}(\{\xi\},k,\omega)\right\vert^2}
\, ,
\end{equation}
where the two-point vertex functions appearing on the right hand
side in the above expression have to be calculated within
perturbation expansion. They are obtained by collecting all
1-particle irreducible Feynman graphs with corresponding external
legs. A closer examination of the loop expansion reveals
\cite{fomo02} that the dynamic response vertex functions
$\mathring{\Gamma}_{\psi\tilde{\psi}^+}$ and
$\mathring{\Gamma}_{\phi_{\|}\tilde{\phi}_{\|}}$ have the general
structure
\begin{eqnarray}\label{gapsipsi}
\mathring{\Gamma}_{\psi\tilde{\psi}^+}(\{\xi\},k,\omega)&=&
\!\! - \,\, \mbox{i}\omega\mathring{\Omega}_{\psi\tilde{\psi}^+}(\{\xi\},k,\omega) \nonumber \\
\!\!&+&\!\!\mathring{\Gamma}_{\psi\psi^+}(\{\xi\},k)
\mathring{\Gamma}^{(d)}_{\psi\tilde{\psi}^+}(\{\xi\},k,\omega) \, ,
\end{eqnarray}
\begin{eqnarray}\label{gaphiphi}
\mathring{\Gamma}_{\phi_{\|}\tilde{\phi}_{\|}}(\{\xi\},k,\omega)&=&
\!\! -
 \,\, \mbox{i}\omega\mathring{\Omega}_{\phi_{\|}\tilde{\phi}_{\|}}(\{\xi\},k,\omega)
\nonumber \\
\!\!&+&\!\!\mathring{\Gamma}_{\phi_{\|}\phi_{\|}}(\{\xi\},k)\mathring{\Gamma}_\|
\, ,
\end{eqnarray}
where $\mathring{\Gamma}_{\psi\psi^+}(\{\xi\},k)$ and
$\mathring{\Gamma}_{\phi_{\|}\phi_{\|}}(\{\xi\},k)$ are the well known static
two point vertex functions of the bicritical GLW-model with a complex OP. We want to
remark that the static vertex functions in (\ref{gapsipsi}) and (\ref{gaphiphi})
are related by
\begin{eqnarray}\label{gastatrel1}
\mathring{\Gamma}_{\psi\psi^+}(\{\xi\},k)=
\frac{1}{2}\mathring{\Gamma}_{\perp\perp}^{(2,0)}(\{\xi\},k)
\end{eqnarray}
and
\begin{eqnarray}\label{gastatrel2}
\mathring{\Gamma}_{\phi_{\|}\phi_{\|}}(\{\xi\},k)=
\mathring{\Gamma}_{\|\|}^{(2,0)}(\{\xi\},k)
\end{eqnarray}
to the static vertex functions introduced in papers I--III for the
 model with real OPs.
Thus the  correlation lengths $\xi_\perp$ and $\xi_\|$ are now
determined by
\begin{equation}\label{xiperp}
\xi_\perp^2(\{\mathring{r}\},\{\mathring{u}\})=
\frac{\partial\ln\mathring{\Gamma}_{\psi\psi^+}(k,\{\mathring{r}\},\{\mathring{u}\})}{\partial
k^2}\Bigg\vert_{k=0} \, ,
\end{equation}
\begin{equation}\label{xipara}
\xi_\|^2(\{\mathring{r}\},\{\mathring{u}\})=
\frac{\partial\ln\mathring{\Gamma}_{\phi_{\|}\phi_{\|}}(k,\{\mathring{r}\},\{\mathring{u}\})}{\partial
k^2}\Bigg\vert_{k=0} \, .
\end{equation}
$\mathring{\Omega}_{\psi\tilde{\psi}^+}$,
$\mathring{\Gamma}^{(d)}_{\psi\tilde{\psi}^+}$ and
$\mathring{\Omega}_{\phi_{\|}\tilde{\phi}_{\|}}$ are purely dynamic functions.
The explicit expressions of these functions are given in appendix \ref{appop}, Eqs. (\ref{ompsipsi}) - (\ref{omphiphi}).
They determine also the dynamic vertex functions
$\mathring{\Gamma}_{\tilde{\psi}\tilde{\psi}^+}$ and
$\mathring{\Gamma}_{\tilde{\phi}_{\|}\tilde{\phi}_{\|}}$ in (\ref{cpsipsivert}) and
(\ref{cphiphivert}). A
proper rearrangement of the perturbative contributions shows that the
relations
\begin{equation}\label{gapsitilderel}
\mathring{\Gamma}_{\tilde{\psi}\tilde{\psi}^+}(\{\xi\},k,\omega)=-2
\Re\left[\mathring{\Omega}_{\psi\tilde{\psi}^+}(\{\xi\},k,\omega)
\mathring{\Gamma}^{(d)}_{\psi\tilde{\psi}^+}(\{\xi\},k,\omega)\right]
\, ,
\end{equation}
\begin{equation}\label{gaphitilderel}
\mathring{\Gamma}_{\tilde{\phi}_{\|}\tilde{\phi}_{\|}}(\{\xi\},k,\omega)=-2\Gamma_\|
\Re\left[\mathring{\Omega}_{\phi_{\|}\tilde{\phi}_{\|}}(\{\xi\},k,\omega)\right]
\end{equation}
hold.  $\Re[.]$ is the real part of the expression in the brackets.

Analogous to (\ref{cpsipsixt}) and (\ref{cphiphixt}) the Fourier transformed dynamic
correlation function of the secondary density is introduced as
\begin{equation}\label{cmimjxt}
\mathring{C}_{mm}(\{\xi\},k,\omega)=\!\! \int\!\! d^dx\!\! \int\!\! dt
e^{-ikx+i\omega t} \langle m_0(x,t)m_0(0,0)\rangle_c
\end{equation}
The connection to the dynamic vertex functions is analogous to
the case of the OP Eqs. (\ref{cpsipsivert} and
(\ref{cphiphivert}):
\begin{equation}\label{cmmvert}
\mathring{C}_{mm}(\{\xi\},k,\omega)=
-\frac{\mathring{\Gamma}_{\tilde{m}\tilde{m}}(\{\xi\},k,\omega)}
{\left\vert\mathring{\Gamma}_{m\tilde{m}}(\{\xi\},k,\omega)\right\vert^2}
\, .
\end{equation}
The dynamic response vertex function of the secondary density has the
general structure
\begin{eqnarray}\label{gamm}
\mathring{\Gamma}_{m\tilde{m}}(\{\xi\},k,\omega)\!\!&=&\!\!
-\mbox{i}\omega\mathring{\Omega}_{m\tilde{m}}(\{\xi\},k,\omega)
\nonumber \\
\!\!&+&\!\!\mathring{\Gamma}_{mm}(\{\xi\},k)
\mathring{\Gamma}^{(d)}_{m\tilde{m}}(\{\xi\},k,\omega)
\end{eqnarray}
where $\mathring{\Gamma}_{mm}(\{\xi\},k)$ is the static two point
vertex function calculated with the extended static functional
(\ref{statfunc}) which already has been introduced in paper III. A
relation corresponding to (\ref{gapsitilderel}) holds also for the
dynamic vertex function of the secondary density. We have
\begin{equation}\label{gamtilderel}
\mathring{\Gamma}_{\tilde{m}\tilde{m}}(\{\xi\},k,\omega)=-2
\Re\left[\mathring{\Omega}_{m\tilde{m}}(\{\xi\},k,\omega)
\mathring{\Gamma}^{(d)}_{m\tilde{m}}(\{\xi\},k,\omega)\right] \, .
\end{equation}

\section{Renormalization}  \label{renorm}

\subsection{Static renormalization}

The renormalization of the  GLW-functional (\ref{hbiglw}) has been
extensively discussed in paper I. The only difference in the present paper is that
we now have to renormalize the complex OP $\psi_0$ instead of the
real vector OP $\vec{\phi}_{\perp 0}$. We introduce the following renormalization factor
\begin{equation}\label{phim}
\psi_0=Z_\psi^{1/2}\psi \ , \qquad  \psi_0^+=Z_\psi^{1/2}\psi^+
\end{equation}
where $Z_\psi$ is a real quantity and identical to $Z_{\phi_\perp}$ in paper I taken
at $n_\perp=2$ and $n_\|=1$. This means
\begin{equation}\label{zpsiphi}
Z_\psi=Z_{\phi_\perp}\Big\vert_{n_\perp=2 \atop n_\|=1} \, .
\end{equation}
The renormalization of the parameters $\mathring{r}_\perp$, $\mathring{r}_\|$
and the couplings $\mathring{u}_\perp$, $\mathring{u}_\|$, $\mathring{u}_\times$
appearing in (\ref{hbiglw}) is given in paper I (see Eqs. (16), (17) and (5)-(7)).
In all relations one has to replace $Z_{\phi_\perp}$ by $Z_\psi$. All renormalization
factors remain valid if one sets $n_\perp=2$  and $n_\|=1$. This is also true
for the $Z$-matrix $\mbox{\boldmath$Z$}_{\phi^2}$ introduced in Eq.(10) of paper I
and the additive renormalization $\mbox{\boldmath$A$}(\{u\})$ defined in Eq. (15)
of paper I.

The renormalization of the parameters in the extended static functional (\ref{statfunc})
has been presented in paper III. As in the case of the bicritical GLW-model all $Z$-factors
and relations between them remain valid if $Z_{\phi_\perp}$ therein is replaced by
$Z_\psi$, and  if one sets $n_\perp=2$ and $n_\|=1$ in explicit expressions.

\subsection{Dynamic renormalization}

Within dynamics auxiliary densities $\tilde{\psi}_0$,
$\tilde{\phi}_{\| 0}$ and $\tilde{m}_0$ corresponding to the two OPs
and the secondary density are introduced \cite{bauja76}. Instead of
renormalization conditions we use the minimal subtraction scheme
\cite{schloms} as in the preceding papers II and III. The auxiliary
density of the complex OP is multiplicatively renormalizable by
introducing complex $Z$-factors:
\begin{equation}\label{psit}
\tilde{\psi}_0=Z_{\tilde{\psi}}^{1/2}\tilde{\psi} \ , \qquad
\tilde{\psi}_0^+=Z_{\tilde{\psi}^{+}}^{1/2}\tilde{\psi}^+
\end{equation}
The complex renormalization factors in (\ref{psit}) fulfill the
relation $Z_{\tilde{\psi^{+}}}=Z_{\tilde{\psi}}^+$. For the
auxiliary densities of the single-component real OP and the
secondary density the corresponding renormalization factors are
introduced:
\begin{equation}\label{phimt}
\tilde{\phi}_{\| 0}=Z_{\tilde{\phi}_{\|}}^{1/2}\tilde{\phi}_{\|} \ ,
\qquad \tilde{m}_0=Z_{\tilde{m}}\tilde{m} \, ,
\end{equation}
where $Z_{\tilde{\phi}_{\|}}$ and $Z_{\tilde{m}}$ are real. Within
the minimal subtraction scheme the $Z$-factors of the auxiliary
densities of the non-conserved OPs $Z_{\tilde{\psi}^{+}}$ and
$Z_{\tilde{\phi}_{\|}}$ are determined by the $\varepsilon$-poles of
the functions $\mathring{\Omega}_{\psi\tilde{\psi}^+}$ and
$\mathring{\Omega}_{\phi_\|\tilde{\phi}_\|}$ introduced in
(\ref{gapsipsi}) and (\ref{gaphiphi}). The corresponding function of
the conserved secondary density $\mathring{\Omega}_{m\tilde{m}}$ in
(\ref{gamm}) does not contain new poles. Therefore one has
\begin{equation}\label{zmtilde}
Z_{\tilde{m}}=Z_m^{-1}
\end{equation}
where $Z_m$ has been introduced in Eq. (30) in paper III.

The kinetic coefficients renormalize as
\begin{equation}\label{kincoren}
\mathring{\Gamma}_\perp=Z_{\Gamma_\perp}\Gamma_\perp \ , \quad
\mathring{\Gamma}_\|=Z_{\Gamma_\|}\Gamma_\| \ , \quad
\mathring{\lambda}=Z_{\lambda}\lambda \, .
\end{equation}
The renormalization of the complex kinetic coefficient $\Gamma_\perp$ in (\ref{kincoren})
leads to a complex $Z_{\Gamma_\perp}$, while the other two renormalization factors
in (\ref{kincoren}) are real valued. $Z_{\Gamma_\perp}$ can be separated into
\begin{equation}\label{zgamperpfak}
Z_{\Gamma_\perp}=Z_\psi^{1/2}Z_{\tilde{\psi}\!^+}^{-1/2}Z_{\Gamma_\perp}^{(d)}
\, ,
\end{equation}
where $Z_{\Gamma_\perp}^{(d)}$ contains the singular contributions of the
dynamic function $\mathring{\Gamma}^{(d)}_{\psi\tilde{\psi}^+}$, appearing in
(\ref{gapsipsi}).

The dynamic equation (\ref{dphipadt}) for the OP $\phi_\|$ contains no mode
coupling term. As a consequence only the kinetic coefficient $\mathring{\Gamma}_\|$
appears in the dynamic vertex function (\ref{gaphiphi}) instead of a function
$\mathring{\Gamma}^{(d)}_{\phi_\|\tilde{\phi}_\|}$. Therefore $Z_{\Gamma_\|}^{(d)}=1$
and we can write
\begin{equation}\label{zgamfak}
Z_{\Gamma_\|}=Z_{\phi_\|}^{1/2}Z_{\tilde{\phi}_{\|}}^{-1/2} \ .
\end{equation}
Using Eq.(\ref{zmtilde}) the kinetic coefficient of the secondary density renormalizes as
\begin{equation}\label{zlamfak}
Z_\lambda=Z_m^2Z_\lambda^{(d)}
\end{equation}
where $Z_\lambda^{(d)}$ contains only the poles of the $k^2$ derivative of
$\mathring{\Gamma}^{(d)}_{m\tilde{m}}$ taken at zero frequency and wave vector modulus.

The mode coupling coefficient needs no independent renormalization, so we simply have
\begin{equation}\label{giren}
\mathring{g}=\kappa^{\varepsilon/2}Z_mgA_d^{-1/2} \, .
\end{equation}
The geometric factor $A_d$ \cite{dohm85} already used in the static renormalization has been given in paper I Eq. (8).

\section{Renormalization Group functions}  \label{rngfunctions}

In order to obtain the temperature dependence of the model parameters, as well as the asymptotic
dynamic exponents, the RG functions, which are usually denoted as $\zeta$- and
$\beta$-functions have to be introduced.

\subsection{General definitions}\label{generaldef}

In order to simplify the general handling of the RG functions we
will use the uniform definition
\begin{equation}\label{zetafu}
\zeta_{a_i}(\{\alpha_j\})=\frac{d\ln Z_{a_i}^{-1}(\{\alpha_j\})}{d\ln\kappa}
\end{equation}
for all $\zeta$-functions in statics and dynamics. The derivative is taken at fixed bare parameters. $\{\alpha_j\}$
denotes the set of static and dynamic model parameters which include
the static couplings $\{u\}$ and $\{\gamma\}$, the mode coupling
$g$, and all kinetic coefficients
$\Gamma_\perp,\Gamma_\perp^+,\Gamma_\|,\lambda$. The
$\zeta$-function $\zeta_{a_i}$ is calculated from the
renormalization factor $Z_{a_i}$ introduced in the previous section.
Thus $a_i$ may denote a model parameter from the set $\{\alpha_j\}$,
a density $\phi_\perp$, $\phi_\|$, $m$, or a composite operator
$\phi_\perp^2$, $\phi_\|^2$.  The approach of the model parameters
$\alpha_i(l)$ to their FP values in the vicinity of the
multicritical point is determined by the flow equations with the flow parameter $l$
\begin{equation}\label{floweq}
l\frac{d\alpha_i(l)}{dl}=\beta_{\alpha_i}(\{\alpha_j(l)\})
\end{equation}
with $\beta$-functions
\begin{equation}\label{betafu}
\beta_{\alpha_i}(\{\alpha_j(l)\})=\alpha_i(l)\Big(-c_i+\zeta_{\alpha_i}(\{\alpha_j(l)\})\Big)
\end{equation}
$c_i$ is the naive dimension of the corresponding parameter $\alpha_i$ obtained by power counting. For the static
couplings $u_\perp$, $u_\times$ or $u_\|$ the naive
dimension $c_i$ is equal to $\varepsilon$, while for $\gamma_\perp$ or $\gamma_\|$ and
the mode coupling $g$ it is $\varepsilon/2$ respectively.
All kinetic coefficients, these are $\Gamma_\perp$, $\Gamma_\perp^+$, $\Gamma_\|$ and $\lambda$, are dimensionless
quantities, which means $c_i=0$.

The flow equations (\ref{floweq}) have fixed points at the zeros of the $\beta$-functions. The FP
values of the model parameters $\{\alpha_j^\star\}$ are defined by the equations
\begin{equation}\label{fixpointeq}
\beta_{\alpha_i}(\{\alpha_j^\star\})=0 \, .
\end{equation}
The FP is stable if all eigenvalues of the matrix
$\partial\beta_{\alpha_i}/\partial\alpha_k$ are positive or
possesses positive real parts. Starting at values
$\{\alpha_j(l_0)\}$ at an initial flow parameter value $l_0$, the
flow equations can be solved numerically. The asymptotic critical
values of the parameters are obtained in the limit $l\to 0$. If a
stable FP is present the flow of the parameters has the property
\begin{equation}\label{flowfix}
\lim_{l\to 0}\{\alpha_j(l)\}=\{\alpha_j^\star\} \, .
\end{equation}
A set of FP values $\{\alpha_j^\star\}$ determines all static and
dynamic exponents. The static relations between $\zeta$-functions
and critical exponents have been extensively discussed in papers I
and III. The dynamic exponents are related by
\begin{equation}\label{dynexpdef}
z_{\phi_\perp}=2+\zeta_{\Gamma_\perp^\prime}^{\star} \ , \quad z_{\phi_\|}=2+\zeta_{\Gamma_\|}^\star \ , \quad
z_m=2+\zeta_\lambda^\star
\end{equation}
to the dynamic $\zeta$-functions (see \cite{review}). In
(\ref{dynexpdef}) the short notation
$\zeta_{\alpha_i}^\star\equiv\zeta_{\alpha_i}(\{\alpha_j^\star\})$
has been introduced. In the non-asymptotic background region
effective dynamic exponents are defined as
\begin{eqnarray}
\label{zperpeff}
z_\perp^{(eff)}(l)&=&2+\zeta_{\Gamma_\perp^\prime}\big(\{\alpha_j(l)\}\big) \ , \\
\label{zparaeff}
z_\|^{(eff)}(l)&=&2+\zeta_{\Gamma_\|}\big(\{\alpha_j(l)\big) \ , \\
\label{zmeff}
z_m^{(eff)}(l)&=&2+\zeta_\lambda\big(\{\alpha_j(l)\big) \ .
\end{eqnarray}
where the flow of the parameters is inserted into the
$\zeta$-functions instead of the FP values. The effective exponents
depend on the flow parameter, or reduced temperature accordingly.
Relation (\ref{flowfix}) makes sure that the effective exponents
turn into the asymptotic exponents in the critical limit, that is
\begin{equation}\label{zfix}
\lim_{l\to 0} z_k^{(eff)}(l)=z_k \qquad \mbox{with} \qquad k=\perp, \|, m
\end{equation}

\subsection{Time scale ratios and mode coupling parameters}

It is convenient to introduce ratios of the kinetic coefficients or mode couplings,
which may have finite FP
values. The following ratios will be used in the subsequent sections:
\begin{itemize}
\item[(i)] The time scale ratios between the order parameters and the secondary density
\begin{equation}\label{w12}
w_\perp\equiv\frac{\Gamma_\perp}{\lambda} \ , \qquad
w_\|\equiv\frac{\Gamma_\|}{\lambda} \, .
\end{equation}
From this we may also define the ratio between kinetic coefficients of the two order parameters
\begin{equation}\label{v}
v\equiv\frac{\Gamma_\|}{\Gamma_\perp}=\frac{w_\|}{w_\perp}
\end{equation}
which already previously has been used in the bicritical model A and model C. Note
that in contrast
to the two models mentioned, $w_\perp$ and $v$ are now complex quantities. The ratios in
Eqs.(\ref{w12}) and (\ref{v}) are of course not independent as shown by the
equality in (\ref{v}). The structure of the dynamic $\zeta$-functions presented subsequently
further implies the introduction of the complex ratio
\begin{equation}\label{vperp}
v_\perp\equiv\frac{\Gamma_\perp}{\Gamma_\perp^+}=\frac{w_\perp}{w_\perp^+}=\frac{v^+}{v}
\, .
\end{equation}

\item[(ii)] The mode coupling parameter
\begin{equation}\label{F}
F\equiv\frac{g}{\lambda} \, .
\end{equation}
The above ratio does not necessarily have a finite FP value. Thus it may be
more appropriate to use the ratio
\begin{equation}\label{fd}
f_\perp\equiv\frac{g}{\sqrt{\Gamma_\perp^{\prime}\lambda}}=\frac{F}{\sqrt{w_\perp^\prime}}
\end{equation}
in several cases, especially in the discussion of the flow equations and the fixed points.
\end{itemize}
The flow equations for the ratios defined above can be found from the $\zeta$- and $\beta$-functions
introduced in the previous subsection. From the definition of the parameters in (\ref{w12}),
(\ref{fd}) and the renormalization (\ref{kincoren}) and
(\ref{giren}) we obtain together with (\ref{zetafu}) the flow equations
\begin{eqnarray}
\label{dw1dl}
l\frac{d w_\perp}{dl}&=&w_\perp\left(\zeta_{\Gamma_\perp}-\zeta_\lambda\right) \ , \\
\label{dw2dl}
l\frac{d w_\|}{dl}&=&w_\|\left(\zeta_{\Gamma_\|}-\zeta_\lambda\right) \ , \\
\label{df1dl}
l\frac{d f_\perp}{dl}&=&-\frac{f_\perp}{2}\left(\varepsilon+\zeta_\lambda-2\zeta_m
+\Re\left[\frac{w_\perp}{w_\perp^\prime}\ \zeta_{\Gamma_\perp}\right]\right) .
\end{eqnarray}
From (\ref{dw1dl}) and (\ref{dw2dl}) follows immediately the flow equation for the ratio
\begin{equation}\label{dvdl}
l\frac{d v}{dl}=v\left(\zeta_{\Gamma_\|}-\zeta_{\Gamma_\perp}\right)\ , \\
\end{equation}
which has been defined in (\ref{v}).

The remaining task is to calculate the explicit expressions of
the dynamic functions $\zeta_{\Gamma_\perp}$, $\zeta_{\Gamma_\|}$ and $\zeta_{\lambda}$ in
two loop order.

\section{Dynamic RG-functions in two loop order} \label{twoloop}

The perturbation expansion of the dynamic vertex functions and the
structures therein are outlined in detail in appendix \ref{appop}.
The outcoming expressions  for the  dynamic
renormalization factors in two loop order are presented in appendix
\ref{appz}. With these expressions at hand we are in the position to
obtain explicit two loop expressions for the RG $\zeta$-functions as
expressed in the following.

\subsection{Dynamic $\zeta$-functions of the OPs}

Relation (\ref{zgamfak}) between the Z-factors implies the relations between the
corresponding $\zeta$-functions
\begin{equation}\label{zetagperel}
\zeta_{\Gamma_\perp}=\zeta_{\Gamma_\perp}^{(d)}-\frac{1}{2}\zeta_{\tilde{\psi}\!^+}
+\frac{1}{2}\zeta_\psi \, ,
\end{equation}
\begin{equation}\label{zetagparel}
\zeta_{\Gamma_\|}=-\frac{1}{2}\zeta_{\tilde{\phi}_\|}
+\frac{1}{2}\zeta_{\phi_\|} \, .
\end{equation}
The static $\zeta$-functions $\zeta_\psi=\zeta_{\phi_\perp}$ has
been presented Eqs.(20) in paper I. Inserting (\ref{zpsit}) and
(\ref{zgammad}) into (\ref{zetafu}) and (\ref{zetagperel}) we obtain
the dynamic $\zeta$-function for the kinetic coefficient of the
perpendicular components as
\begin{eqnarray}\label{zetagammape}
\zeta_{\Gamma_{\perp}}=\frac{D_\perp^2}{w_\perp(1+w_\perp)}
-\frac{2}{3}\frac{u_\perp D_\perp}{w_\perp(1+w_\perp)}\ A_\perp \nonumber \\
-\frac{1}{2}\frac{D_\perp^2}{w_\perp^2(1+w_\perp)^2}\ B_\perp  \nonumber \\
-\frac{1}{2}\frac{\gamma_\|D_\perp}{1+w_\perp}\left(\frac{u_\times}{3}
+\frac{1}{2}\frac{\gamma_\|D_\perp}{1+w_\perp}\right)X_\perp  \nonumber \\
+\zeta_{\Gamma_{\perp}}^{(A)}\big(\{u\},v_\perp,v\big)
\end{eqnarray}
where we have introduced the coupling
\begin{equation}\label{dperp}
D_\perp\equiv w_\perp\gamma_\perp-\mbox{i}F \, .
\end{equation}
The functions $A_\perp$, $B_\perp$ and $X_\perp$ are defined as
\begin{equation}\label{aperp}
A_\perp\equiv w_\perp\gamma_\perp(1-x_1L_1)+\mbox{i}Fx_-x_1L_1-D_\perp L_0
\end{equation}
\begin{eqnarray}\label{bperp}
B_\perp\!\!\!&\equiv&\!\!\!w_\perp^2\gamma_\perp^2(1-2x_1L_1)
+F^2(2x_-L_1+L_R) \nonumber \\
\!\!\!&&\!\!\!+2w_\perp\gamma_\perp\mbox{i}F(1+2x_-x_1L_1)-2L_0D_\perp^2 \nonumber \\
\!\!\!&-&\!\!\!\frac{D_\perp^2}{1+w_\perp}\Bigg(w_\perp+(1+2w_\perp)
\ln\frac{(1+w_\perp)^2}{1+2w_\perp}\Bigg)
\end{eqnarray}
\begin{equation}\label{xperp}
X_\perp\equiv 1+\ln\frac{2v}{1+v}-\left(1+\frac{2}{v}\right)\ln\frac{2(1+v)}{2+v}
\end{equation}
with
\begin{eqnarray}\label{lr}
L_R\equiv\Big[x_++v_\perp+x_+^2(x_+^2+2v_\perp^2)\Big]
\frac{L_1}{x_+} -3v_\perp \, .
\end{eqnarray}
We have used the following definitions in the above
expressions:
\begin{equation}\label{x1xpm}
x_\pm\equiv 1\pm v_\perp \ , \qquad x_1\equiv 2+v_\perp \, ,
\end{equation}
\begin{equation}\label{l0l1}
L_0\equiv 2\ln\frac{2}{1+\frac{1}{v_\perp}}\ , \qquad
L_1\equiv\ln\frac{\left(1+\frac{1}{v_\perp}\right)^2}
{1+2\frac{1}{v_\perp}} \, .
\end{equation}
$\zeta_{\Gamma_{\perp}}^{(A)}\big(\{u\},v_{\perp},v\big)$ is the
$\zeta$-function of the kinetic coefficient of the perpendicular components
in the bicritical model A, but now with a complex kinetic coefficient $\Gamma_\perp$.
It reads in two loop order
\begin{eqnarray}\label{zetagaape}
\zeta_{\Gamma_{\perp}}^{(A)}\big(\{u\},v_{\perp},v\big)=
\frac{u_\perp^2}{9}\left(L_0+x_1L_1-\frac{1}{2}\right) \nonumber \\
+\frac{u_\times^2}{36}\left(L_\perp^{(\times)}-\frac{1}{2}\right)
\end{eqnarray}
with
\begin{equation}\label{lperp}
L_\perp^{(\times)}\equiv\ln\frac{(1+v)^2}{v(2+v)}+\frac{2}{v}\ln\frac{2(1+v)}{2+v}
\, .
\end{equation}
The dynamic $\zeta$-function of the parallel component is obtained
by inserting Eq.(21) of paper I and (\ref{zgammapa}) into
(\ref{zetafu}) and (\ref{zetagparel}). The result is
\begin{eqnarray}\label{zetagammapa}
\zeta_{\Gamma_{\|}}=\frac{w_\|\gamma_\|^2}{1+w_\|}
-\frac{1}{2}\frac{w_\|\gamma_\|}{1+w_\|}\Bigg[u_\| \gamma_\|\left(1-3\ln\frac{4}{3}\right)
\nonumber \\
+\frac{w_\|\gamma_\|^3}{1+w_\|}\Bigg(\frac{1}{2}\left(1-9\ln\frac{4}{3}\right)
- \frac{w_\|}{1+w_\|} \nonumber \\
- \frac{1+2w_\|}{1+w_\|}\ln\frac{(1+w_\|)^2}{1+2w_\|}\Bigg)  \nonumber \\
+\left(\frac{2}{3}u_\times{+}\frac{w_\|\gamma_\|}{1{+}w_\|}\gamma_\perp\right)
\Re\Big[\frac{T_1}{w_\perp^\prime}\Big]-\frac{\gamma_\|F}{2w_\perp^\prime(1{+}w_\|)}
\Im\Big[\frac{T_2}{w_\perp^\prime}\Big]\Bigg]  \nonumber \\
+\zeta_{\Gamma_{\|}}^{(A)}\big(\{u\},v_\perp,v\big)  \, . \nonumber \\
\end{eqnarray}
The functions $T_1$ and $T_2$ are defined as
\begin{eqnarray}\label{T1}
T_1\equiv&&\!\!\!\!D_\perp\Bigg[1+\ln\frac{1+\frac{1}{v_\perp}}{1+v}  \nonumber\\
&&\!\!\!\!-\left(v{+}\frac{1}{v_\perp}(1{+}v)\right)
\ln\frac{(1{+}v)\left(1{+}\frac{1}{v_\perp}\right)}{v{+}\frac{1}{v_\perp}(1{+}v)}\Bigg]
\, ,
\end{eqnarray}
\begin{eqnarray}\label{T2}
T_2\equiv&&\!\!\!\!w_\perp^+D_\perp\Bigg[(1+v_\perp)v
-\ln\frac{1+\frac{1}{v_\perp}}{1+v}  \nonumber\\
&&\!\!\!\!-\left(v{+}\frac{1}{v_\perp}(1{+}v)\right)\big(v{+}v_\perp(1{+}v)\big)
\nonumber  \\
&&\!\!\!\!\times
\ln\frac{(1{+}v)\left(1{+}\frac{1}{v_\perp}\right)}{v{+}\frac{1}{v_\perp}(1{+}v)}\Bigg]
\, ,
\end{eqnarray}
$\zeta_{\Gamma_{\|}}^{(A)}\big(\{u\},v_{\perp},v\big)$ is the
$\zeta$-function of the kinetic coefficient of the parallel component
in the bicritical model A. With a complex $\Gamma_\perp$ it reads
\begin{eqnarray}\label{zetagaapa}
\zeta_{\Gamma_{\|}}^{(A)}\big(\{u\},v_{\perp},v\big)=
\frac{u_\|^2}{4}\left(\ln\frac{4}{3}-\frac{1}{6}\right)
+\frac{u_\times^2}{18}\left(L_\|^{(\times)}-\frac{1}{2}\right)   \nonumber \\
\end{eqnarray}
with
\begin{eqnarray}\label{TA}
L_\|^{(\times)}\equiv\ln\frac{(1{+}v)\left(\frac{1}{v_\perp}{+}v\right)}
{v{+}\frac{1}{v_\perp}(1{+}v)}
+vv_\perp\ln\frac{\left(1{+}\frac{1}{v_\perp}\right)\left(\frac{1}{v_\perp}{+}v\right)}
{v{+}\frac{1}{v_\perp}(1{+}v)}   \nonumber \\
+v\ln\frac{\left(1{+}\frac{1}{v_\perp}\right)(1{+}v)}{v{+}\frac{1}{v_\perp}(1{+}v)}
\, .
\nonumber \\
\end{eqnarray}

\subsection{Dynamic $\zeta$-functions of the secondary density}

With relation (\ref{zlamfak}) we can separate the static contributions
to the $\zeta$-function $\zeta_\lambda$. Thus we have
\begin{equation}\label{zetasep}
\zeta_\lambda=2\zeta_m+\zeta_\lambda^{(d)}
\end{equation}
By separating the static from the dynamic parts in the $\zeta$-functions one
can take advantage of the general structures appearing in the purely dynamic
$\zeta$-function $\zeta_\lambda^{(d)}$ as well as in the static $\zeta$-function
$\zeta_m$. Inserting $n_\perp=2$ and $n_\|=1$ into relation (40) in paper III $\zeta_m$ can be written as
\begin{equation}\label{zetam}
\zeta_m=\frac{1}{2}\gamma_\perp^2+\frac{1}{4}\gamma_\|^2
\end{equation}
which is valid up to two loop order. From the diagrammatic structure of the dynamic
perturbation theory follows
\begin{equation}\label{zetalad}
\zeta_\lambda^{(d)}=-\frac{f_\perp^2}{2}\Big(1+{\cal Q}\Big) \, .
\end{equation}

The real function ${\cal Q}$ contains all higher order contributions beginning
with
two loop order. Setting ${\cal Q}=0$ in (\ref{zetalad}) reproduces
the one loop expressions of this function.
The function ${\cal Q}$ in the dynamic $\zeta$-function of the secondary
density (\ref{zetalad}) has the structure
\begin{equation}\label{qfunc}
{\cal Q}=\frac{1}{2}\Re[X_2]
\end{equation}
from which immediately follows that it is a real quantity. $X_2$ reads
\begin{eqnarray}\label{x2}
X_2=&&\!\!\!\frac{D_\perp}{w_\perp^\prime (1+w_\perp)}
\Bigg[D_\perp\left(\frac{1}{2}+\ln\frac{1+w_\perp}{1+w_\perp^+}\right)  \\
&&\!\!\!\!\!+D_\perp^+(1+w_\perp)
-\left(W_\perp^{(m)}\gamma_\perp+w_\perp\mbox{i}F\right)
W_\perp^{(m)}L_\perp^{(m)}\Bigg]   \nonumber
\end{eqnarray}
where we have introduced the definitions
\begin{equation}\label{lmk}
L_\perp^{(m)}=\ln\left(1+\frac{1}{W_\perp^{(m)}}\right) \, ,
\end{equation}
\begin{equation}\label{wmk}
W_\perp^{(m)}=w_\perp+w_\perp^++w_\perp w_\perp^+\, .
\end{equation}
Note that $X_2$  coincides  with the corresponding function in
model F in \cite{dohm85d, review}.

\section{Critical behavior in one loop order}\label{onelooporder}

Although the one loop critical behavior of the considered system has already been discussed in
\cite{dohmjanssen77} we want to summarize the results in order to compare it with the considerably differing
results of the two loop calculation. In one loop order the $\zeta$-functions (\ref{zetagammape}),
(\ref{zetagammapa}) and (\ref{zetalad}) reduce to
\begin{eqnarray}\label{zetagammape1L}
\zeta_{\Gamma_{\perp}}=\frac{D_\perp^2}{w_\perp(1+w_\perp)} \ , \quad
\zeta_{\Gamma_{\|}}=\frac{w_\|\gamma_\|^2}{1+w_\|} \ , \quad \zeta_\lambda^{(d)}=-\frac{f_\perp^2}{2} \, .
\nonumber \\
\end{eqnarray}
Inserting (\ref{zetagammape1L}) into the right hand sides of
(\ref{dw1dl})-(\ref{df1dl}) leads to a set of equations in which the
zeros determine the dynamical FPs. The only stable FP is found for
$w_\|^\star=0$ and $w_\perp^{\prime\star}$,
$w_\perp^{\prime\prime\star}$, $f_\perp^\star$ finite. The
corresponding values are presented in Tab. \ref{tab3}.  As a
consequence we have $v^\star=0$. We want to note that the static FP
values of the two loop calculation in paper I and III have been
used. We were interested in the non-asymptotic properties described
by flow equations and since no real FP in statics is reached in two
loop order we had to resum the static $\beta$-functions in order to
get real FP values \cite{partI}.  To each type of static FP (biconical or
Heisenberg) two equivalent static fixed points exist differing in
the signs of $\gamma_\perp$ and $\gamma_\|$ \cite{partIII}. Accordingly four
equivalent dynamic fixed points exist with different signs in
$w_\perp^{\prime\prime}$ and $f_\perp$. They correspond to the
directions of the external fields of the parallel and perpendicular
OP.

The finite value of $w_\perp^\star$ implies the relations
\begin{equation}\label{zetarelfix1L}
\zeta_{\Gamma_\perp}^{\prime\star}=\zeta_\lambda^\star \ , \quad  \zeta_{\Gamma_\perp}^{\prime\prime\star}=0 \ ,
\quad  \varepsilon+\zeta_{\Gamma_\perp}^{\prime\star}+\zeta_\lambda^\star-2\zeta_m^\star=0 \ ,
\end{equation}
which follow from (\ref{dw1dl}) and (\ref{df1dl}). The vanishing
$w_\|^\star$ leads to $\zeta_{\Gamma_\|}^\star=0$ as can immediately
be seen from (\ref{zetagammape1L}). Using the first relation in
(\ref{zetarelfix1L}) and the third one, we obtain
\begin{equation}\label{zetagplam1L}
\zeta_{\Gamma_\perp}^{\prime\star}=\zeta_\lambda^\star =
\frac{1}{2}(2\zeta_m^\star-\varepsilon) \, .
\end{equation}
\begin{figure}[b]
      \centering{
       \epsfig{file=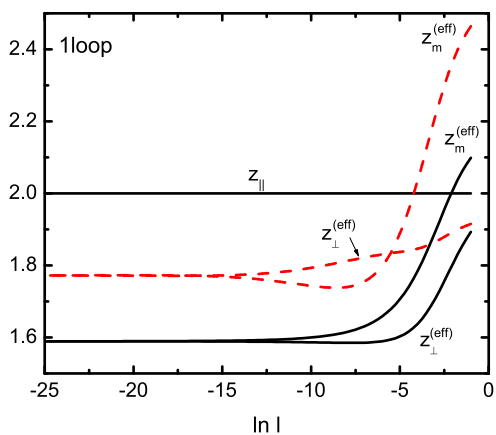,width=8cm,angle=0}}
\caption{ \label{z1loop} Effective dynamic exponents at $d=3$ calculated in
one loop order using the one loop expression for the flow
equations (\ref{dw1dl}), (\ref{df1dl}). The effective exponents are
calculated at the biconical FP (full lines) and at the Heisenberg FP
(dashed line). $z_\|$ is valid for both FPs.}
\end{figure}
\begin{figure}[t]
      \centering{
       \epsfig{file=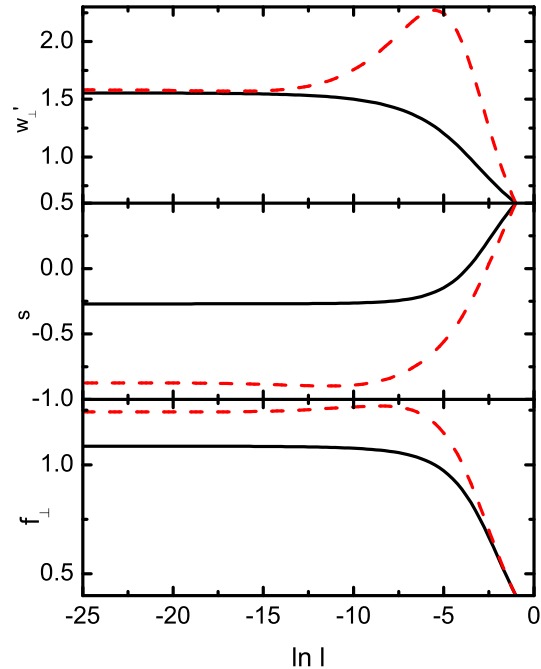,width=8cm,angle=0}}
\caption{ \label{flow1loop} Flow of the parameters $w_\perp^\prime$,
$s=w_\perp^{\prime\prime}/w_\perp^\prime$ and $f_\perp$ in one loop
order at $d=3$. The calculation has been performed for the biconical (solid
lines) and the Heisenberg (dashed lines) FP.}
\end{figure}
Inserting the FP value of the static $\zeta$-function $\zeta_m^\star$ (see relation (105) in paper III)
\begin{equation}\label{zetamstar}
\zeta_m^\star=\frac{\phi}{\nu}-\frac{d}{2}
\end{equation}
into the above equation one has
\begin{equation}\label{zetagplam1Lend}
\zeta_{\Gamma_\perp}^{\prime\star}=\zeta_\lambda^\star =
\frac{\phi}{\nu}-2 \, .
\end{equation}
The dynamic critical exponents (\ref{dynexpdef}) in one loop order
are therefore completely expressed in terms of the static
exponents:
\begin{equation}\label{dynexp1L}
z_\perp\equiv z_{\phi_\perp}=z_m=\frac{\phi}{\nu} \ , \quad z_\|\equiv z_{\phi_\|}=2 \, .
\end{equation}
These static exponents might also be taken from static experiments.
All our numerical calculations are performed in $d=3$ ($\epsilon=1$).
The numerical values of the static exponents $\phi$ and $\nu$ have
been calculated in two loop order in paper I  and are given there in
Tab.III ($\nu=\nu_+$ therein) in two loop order resummed. In one
loop order the two OPs have different dynamic critical exponents.
Scaling is fulfilled only between the perpendicular OP and the
secondary density. The parallel OP behaves like the van Hove model.
This is demonstrated in Fig.\ref{z1loop}, where the effective
exponents defined in (\ref{zperpeff}) - (\ref{zmeff}) have been
calculated by using the flow equations in one loop order. At a flow
parameter about $l\sim e^{-15}$ for both, the biconical FP (solid
lines) and the Heisenberg FP (dashed lines), the asymptotic values
of the dynamic exponents $z_\perp$ and $z_m$ are reached. The
classical value $z_\|=2$, valid for both static fixed points, also
is indicated by a straight line. The corresponding flow is presented
in Fig.\ref{flow1loop}, which proofs that the dynamic exponents in
Fig.\ref{z1loop} have reached their asymptotic behavior because the
dynamic parameters are at their FP values at $\ln l =-25$.
 \begin{table*}[htb]
\centering \tabcolsep=3mm
\begin{tabular}{rrrrrrrrr}
 \hline \hline
 FP & $u_\|^\star$ & $u_\perp^\star$ & $u_\times^\star$ & $\gamma_\|^\star$ & $\gamma_\perp\star$
&  $w^{\prime\star}_\perp$    & $w^{\prime\prime\star}_\perp$ & $f_\perp^\star$
\\ \hline
${\mathcal B}$  & 1.28745 & 1.12769 & 0.30129 & 0.54201 &  -0.17806  &  1.55489  & $\mp$ 0.41958  & $\pm$ 1.08563   \\
${\mathcal B}$  & 1.28745 & 1.12769 & 0.30129 & -0.54201 &  0.17806  &   1.55489  & $\pm$ 0.41958  & $\pm$ 1.08563    \\
${\mathcal H}$ & 1.00156  & 1.00156 & 1.00156 &  0.85179 & -0.42590   &   1.58136 & $\mp$ 1.38256    &$\pm$  1.24264    \\
${\mathcal H}$ & 1.00156  &  1.00156 & 1.00156 &    -0.85179 & 0.42590   &   1.58136 & $\pm$ 1.38256     & $\pm$ 1.24264  \\
\hline \hline
\end{tabular}
\caption{\label{tab3} FP values of  couplings and timescale ratios
for $n_\|=1$, $n_\perp=2$ at $d=3$. ${\mathcal B}$ indicates the biconical,
${\mathcal H}$ the Heisenberg FP. There are always two equivalent
static FPs depending on the signs of the couplings $\gamma$. The FP
values of the static couplings $\{u\}$ and $\{\gamma\}$ are taken
from the resummed two-loop results \cite{partI}, whereas $
w^\star_\perp$ and $f_\perp^\star$ are calculated from the one-loop
$\beta$-functions. $w^\star_\|=v^\star=0$ is valid in all cases.
Corresponding to the two equivalent cases in statics and the sign of
$w^{\prime\prime\star}_\perp$  there are equivalent dynamic FPs with
corresponding signs of the FP value of the mode coupling
$f_\perp^\star$.}
\end{table*}

\section{Limiting behavior of the dynamical $\zeta$-functions in 2-loop order}\label{limitzeta}

The appearance of $\ln v$-terms in the two loop contribution to the
$\zeta_{\Gamma_\perp}$-function, Eq (\ref{zetagammape}),  changes
the discussion of the fixed points considerably compared to the one
loop case. In order to determine the dynamical fixed points of the
current model in two loop order it is necessary to know something
about the limiting behavior of the $\zeta$-functions. For this
reason we will present the $\zeta$-functions in cases where one or
several dynamical parameters go to zero or infinity under definite
conditions. This is necessary because some $\zeta$-functions exhibit
singular behavior under these conditions, which influences the
discussion of possible fixed points. It is anticipated that the
critical exponents defined by the values of the $\zeta$-functions at
the FP are finite and real.

The  auxiliary functions $X_\perp$, $L_\perp^{(\times)}$, $T_1$,
$T_2$ and $L_\|^{(\times)}$, which appear in the $\zeta$-functions
(\ref{zetagammape}) and (\ref{zetagammapa}) behave singularly in
several limits of the parameters. Thus several FP values of the
different parameters can be excluded due to diverging
$\zeta$-functions.  For a summary of the subsequent analysis of the
$\zeta$-function on the time scale ratios see Tab. \ref{table3}.

i) At first we will consider the two functions $X_\perp$ and $L_\perp^{(\times)}$ in (\ref{xperp})
and (\ref{lperp}),
which appear in $\zeta_{\Gamma_\perp}$ and depend on $v$ only. These two functions remain regular if
$v$ grows to infinity. In this case one simply has $X_\perp(v\to\infty)=1$ and $L_\perp^{(\times)}(v\to\infty)=0$.
But for vanishing $v$ both functions evolve a term proportional to $\ln v$. One gets
\begin{equation}\label{xplplim}
X_\perp(v\to 0) = \ln(2v) \, , \qquad L_\perp^{(\times)}(v\to
0)=1-\ln(2v) \, .
\end{equation}
Thus divergent $\ln v$ terms appear in $\zeta_{\Gamma_\perp}(v\to 0)$ independent from the individual
behavior of $w_\perp$ and $w_\|$ because only the ratio $v$ enters the function.

ii) The dynamic $\zeta$-function (\ref{zetagammapa}) of the parallel component contains the three functions
$T_1$, $T_2$ and $L_\|^{(\times)}$ defined in (\ref{T1}), (\ref{T2}) and (\ref{TA}) which contain the ratio $v$.
These functions, and therefore also $\zeta_{\Gamma_\|}$, remain non-divergent for vanishing $v$. One obtains
\begin{eqnarray}\label{limt1t1tav0}
T_1(v\to
0)\!\!\!&=&\!\!\!D_\perp\left(1+\ln\left(1+\frac{1}{v_\perp}\right)-\frac{1}{v_\perp}\ln(1+v_\perp)\right)
\, ,
\nonumber \\
T_2(v\to 0)\!\!\!&=&\!\!\!-w_\perp^\star
D_\perp\left(\ln\left(1+\frac{1}{v_\perp}\right)+\ln(1+v_\perp)\right)
\, ,
\nonumber \\
L_\|^{(\times)}(v\to 0)\!\!\!&=&\!\!\!0 \, .
\end{eqnarray}
But they diverge when $v$ is growing to infinity:
\begin{eqnarray}\label{limt1t1tavinf}
T_1(v\to
\infty)\!\!\!&=&\!\!\!D_\perp\ln\frac{\left(1+\frac{1}{v_\perp}\right)}{v}
\, ,
\nonumber \\
T_2(v\to \infty)\!\!\!&=&\!\!\!-w_\perp^\star
D_\perp\ln\frac{\left(1+\frac{1}{v_\perp}\right)}{v} \, ,
\nonumber \\
L_\|^{(\times)}(v\to
\infty)\!\!\!&=&\!\!\!1+\ln\frac{v}{\left(1+\frac{1}{v_\perp}\right)}
\, .
\end{eqnarray}
In contrast to the case i) the function $\zeta_{\Gamma_\|}$ in
the parallel subspace evolves logarithmic terms $\ln v$ in the
limit $v\to\infty$ and stays finite in the limit $v\to 0$. The above
discussion is also independent of the individual behavior of
$w_\perp$ and $w_\|$ because only the ratio $v_\perp$ stays always
finite and the three functions $T_1$, $T_2$ and $L_\|^{(\times)}$
remain finite even for diverging time scale ratios if their
prefactors are taken into account.
\begin{table}[h,t]
\centering \tabcolsep=5mm
\begin{tabular}{cccc}
 \hline \hline
 Limit & $\zeta_{\Gamma_\perp}$ & $\zeta_{\Gamma_\|}$ & $\zeta_\lambda$  \\ \hline
$v\to 0$ & $\sim\ln v$ & regular & unaffected \\
$v\to\infty$ & regular & $\sim\ln v$ & unaffected \\
$w_\perp\to\infty$ & $\sim\ln w_\perp$ & regular & $\sim w_\perp^{\prime 2}$ \\
$w_\|\to\infty$ & regular & $\sim\ln w_\|$ & unaffected \\
$w_\perp\to 0$  & regular & $\sim\ln v$ & regular \\
$w_\|\to 0$  & $\sim\ln v$ & regular & unaffected \\ \hline
\end{tabular}
\caption{\label{table3} Limiting behavior of the dynamic $\zeta$-functions}
\end{table}

iii) Additional logarithmic singularities may arise in the dynamic $\zeta$-functions if the time scale
ratios $w_\perp$ and $w_\|$ grow individually to infinity independent of the behavior of $v$.
A closer examination of (\ref{zetagammape}) reveals that in the limit $w_\perp\to\infty$
the $\zeta$-function is proportional to
\begin{equation}\label{zgpelimwinf}
\zeta_{\Gamma_\perp}(w_\perp\to\infty) \sim \frac{1}{2}\gamma_\perp^4 \ln\frac{w_\perp}{2}
\end{equation}
independent of the behavior of $v$. Quite analogously the same happens in (\ref{zetagammapa}) when
$w_\|$ grows to infinity. One obtains
\begin{equation}\label{zgpalimwinf}
\zeta_{\Gamma_\|}(w_\|\to\infty) \sim \frac{1}{2}\gamma_\|^4
\ln\frac{w_\|}{2} \, .
\end{equation}
Supposing a finite (different from zero or infinity) FP value
$f_\perp^\star$ for the mode coupling parameter we may conclude the
following concerning the allowed FP values of the remaining
parameters:

a) From i) and ii) follows that $v^\star$ has to be also different from zero or infinity, otherwise $\ln v$
contributions would lead to divergent $\zeta$-functions.

b) From iii) follows that the finite $v^\star$ only can be realized either by $w_\|$ and $w_\perp$ both finite,
or $w_\|$ and $w_\perp$ both going to zero in the same way. The possibility that both time scale ratios are going
to infinity in the same way is excluded because of the $\ln w_\|$ and $\ln w_\perp$ terms appearing in this case.

\section{General asymptotic relations}\label{generalbehavior}

The FP values $\{\alpha_j^\star\}$ of the model parameters are found from the zeros of the $\beta$-functions
in Eqs.(\ref{dw1dl})-(\ref{df1dl}). From the right hand side of the equations one obtains
\begin{eqnarray}
\label{betawperps}
w_\perp^\star\left(\zeta_{\Gamma_\perp}^\star-\zeta_\lambda^\star \right)&=& 0 \ , \\
\label{betawparas}
w_\|^\star\left(\zeta_{\Gamma_\|}^\star-\zeta_\lambda^\star\right)&=& 0 \ , \\
\label{betafperps}
f_\perp^\star\left(\varepsilon+\zeta_\lambda^\star-2\zeta_m^\star
+\Re\left[\frac{w_\perp^\star}{w_\perp^{\prime^\star}}\ \zeta_{\Gamma_\perp}^\star\right]\right)&=& 0 \ .
\end{eqnarray}
A FP which fulfills Eqs.(\ref{betawperps})-(\ref{betafperps}) has to be also a solution of
\begin{equation}\label{betavs}
v^\star\left(\zeta_{\Gamma_\|}^\star-\zeta_{\Gamma_\perp}^\star\right)= 0
\end{equation}
which follows from (\ref{dvdl}).
The $\zeta$-function $\zeta_{\Gamma_\perp}$ for the perpendicular component of
the OP relaxation is a complex function. Separating real and imaginary part leads to
\begin{equation}
\zeta_{\Gamma_\perp}=\zeta^\prime_{\Gamma_\perp}+i\zeta^{\prime\prime}_{\Gamma_\perp}
\, .
\end{equation}
As a consequence also the equations (\ref{betawperps}) for $w_\perp$ and (\ref{betavs}) for $v$ are complex expressions.
The $\zeta$-function for $\Gamma_\perp^\prime$ is
\begin{equation}\label{zetagammaprime}
\zeta_{\Gamma_\perp^\prime}=\zeta^\prime_{\Gamma_\perp}-s\zeta^{\prime\prime}_{\Gamma_\perp}
\end{equation}
with $s$ defined in (\ref{sq}).

We anticipate that in a real physical system definite dynamical exponents exist,
and therefore the dynamic $\zeta$-functions have to be finite at the
stable FP. As already mentioned in subsection \ref{limitzeta}, the
$\zeta$-functions contain $\ln v$ terms requiring a finite FP value
$v^\star$ in order to obtain finite dynamical exponents. Separating
(\ref{betavs}) into real and imaginary part one has
\begin{eqnarray}
\label{realv}
v^{\prime\star}(\zeta^{\prime\star}_{\Gamma_\perp}-\zeta^\star_{\Gamma_\|})
-v^{\prime\prime\star}\zeta^{\prime\prime\star}_{\Gamma_\perp}=0 \, ,               \\
\label{imagv}
v^{\prime\star}\zeta^{\prime\prime\star}_{\Gamma_\perp}+
v^{\prime\prime\star}(\zeta^{\prime\star}_{\Gamma_\perp}-\zeta^\star_{\Gamma_\|})=0
\, .
\end{eqnarray}
From these two equations the FP relations
\begin{equation}\label{zetarelfix1}
\zeta_{\Gamma_\perp}^{\prime\star}=\zeta_{\Gamma_\|}^\star \ , \qquad  \zeta_{\Gamma_\perp}^{\prime\prime\star}=0
\end{equation}
immediately follow. The second equation in (\ref{zetarelfix1}) implies that the dynamical exponent $z_{\phi_\perp}$
in (\ref{dynexpdef}) can be written as
\begin{equation}\label{zphiperp}
z_{\phi_\perp}=2+\zeta_{\Gamma_\perp}^{\prime\star} \, .
\end{equation}
From the first relation in (\ref{zetarelfix1}) follows
\begin{eqnarray} \label{zperppara}
z_{\phi_\perp}=z_{\phi_\|} \ .
\end{eqnarray}
This means that in the case of a finite FP value $v^\star$ scaling between the OPs is valid.
In order to obtain a critical behavior different from model C, the FP value $f_\perp^\star$ of the mode
coupling parameter also has to be finite and different from zero. Then from Eq.(\ref{betafperps}) follows
\begin{equation}\label{finitef}
\varepsilon+\zeta_{\Gamma_\perp}^{\prime\star}+\zeta_\lambda^\star-2\zeta_m^\star=0
\, ,
\end{equation}
where the second relation of (\ref{zetarelfix1}) already has been
used. Inserting (\ref{zetasep}) and (\ref{zetamstar}) into
(\ref{finitef}) one obtains the relation
\begin{equation}\label{zphiperpzmrel}
z_{\phi_\perp}+z_m=2\frac{\phi}{\nu}
\end{equation}
between the exponents. In summary, the condition that both $v^\star$
and $f_\perp^\star$ have to be finite leads to the two relations
(\ref{zperppara}) and (\ref{zphiperpzmrel}) between the exponents.
Further relations are dependent whether the FP values of the time
scale ratios $w_\perp$ and $w_\|$ are finite or zero and lead to the
following cases.

\begin{itemize}
\item[(i)] {\bf Dynamical strong scaling FP:}

In the case that $w_\perp$ and $w_\|$ are finite at the FP, from (\ref{betawperps}) and (\ref{betawparas})
the relation
\begin{eqnarray}
\zeta_{\Gamma_\perp}^{\prime\star}=\zeta_{\Gamma_\|}^{\star}=\zeta_\lambda^\star
\end{eqnarray}
is obtained, where (\ref{zetarelfix1}) already has been used. From (\ref{dynexpdef}) it follows immediately that
the dynamical exponents have to fulfill the relations
\begin{eqnarray}\label{zstrong}
z_{\phi_\perp}=z_{\phi_\|}=z_m\equiv z \, .
\end{eqnarray}
Thus in the case of strong scaling one dynamical exponent $z$ exists only. The exact value of this exponent can be
found by inserting (\ref{zstrong}) into (\ref{zphiperpzmrel}). One obtains
\begin{eqnarray}\label{zstrongvalue}
z=\frac{\phi}{\nu} \, .
\end{eqnarray}

\item[(ii)] {\bf Dynamical weak scaling FP:}

In the case that $w_\perp$ and $w_\|$ are zero with $v$ finite at the FP, Eqs.(\ref{betawperps}) and
(\ref{betawparas}) are trivially fulfilled and no additional relation between the $\zeta$-functions, and dynamical
exponents respectively, arises. As a consequence two dynamical exponents exist. The first one
\begin{equation}\label{zweakop}
z_{\phi_\|}=z_{\phi_\perp}=z_{OP}
\end{equation}
for the OPs, follows from relation (\ref{zperppara}). The second one
\begin{equation}\label{zweaksd}
z_m=2\frac{\phi}{\nu}-z_{OP}
\end{equation}
for the secondary density, is obtained from (\ref{zphiperpzmrel}).
\end{itemize}

A closer examination of the $\beta$-functions (\ref{betawperps}) -
(\ref{betafperps}), also with numerical methods in $d=3$, reveals
that no FP solution can be found where both, $w_\perp$ and $w_\|$,
are finite. Thus the only solution in $d=3$ which remains is
$w_\perp^{\prime\star}=w_\perp^{\prime\prime\star}=w_\|^{\star}=0$
with $v^\star$ and $v_\perp^\star$ finite. This result of course
depends also on the specific numerical values \cite{notefp} of the static FPs
(given in Tab \ref{tab3}) used in the dynamical equations. The
stable FP lies then in the subspace where the time scale ratios
$w_\|$ and $w_\perp$ approach zero in such a way that their ratios
$v$ and $v_\perp$ remain finite and in general complex quantities.
In order to obtain the finite FP values for $v$ and $v_\perp$ the
two loop $\zeta$-functions may be reduced by setting $w_\perp$ and
$w_\|$ equal to zero by keeping their ratios finite. This will be
performed in the following section.

\section{Critical behavior in the asymptotic subspace}\label{asymptotics}

Since the asymmetric couplings $\gamma_\alpha$ always appear together with the time scale ratios
$w_\alpha$ all terms proportional to these couplings drop out in the asymptotic limit where $w_\alpha \to 0$.
It is convenient to introduce the real ratios
\begin{equation}\label{sq}
s\equiv\frac{w_\perp^{\prime\prime}}{w_\perp^\prime}=\frac{\Gamma_\perp^{\prime\prime}}{\Gamma_\perp^\prime}
\ , \qquad
q\equiv\frac{w_\|}{w_\perp^\prime}=\frac{\Gamma_\|}{\Gamma_\perp^\prime}
\, .
\end{equation}
Thus only $s$, $q$ and $f_\perp$ remain as independent dynamical variables.

The ratio $s$ determines the behavior of the
imaginary part of $w_\perp$ with respect to the real part, while the ratio $q$ indicates the behavior of
$w_\|$ with respect to the real part of $w_\perp$. The complex parameters $v_\perp$ and $v$, introduced in (\ref{vperp})
and (\ref{v}), are expressed by $s$ and $q$ as
\begin{equation}\label{vperpv}
v_\perp=\frac{1+is}{1-is} \ , \qquad  v=\frac{q}{1+is}
\end{equation}
in the following expressions.

\subsection{$\zeta$-functions}

We discuss the behavior of the $\zeta$-functions in the limit
$w_\perp \to 0$ and $w_\| \to 0$ for  $s$ and $q$ constant.

 {\bf Case} {\boldmath$s\ne 0:$}

For $w_\perp=0$ and $w_\|=0$ the $\zeta$-function (\ref{zetagammape}), reduces to
\begin{eqnarray}\label{zetagammape00}
\zeta_{\Gamma_{\perp}}^{(as)}\big(\{u\},s,q,f_\perp\big)=-\frac{f_\perp^2}{1+is} \Bigg\{1+\frac{2}{3}u_\perp\Big(L_0(s)
\nonumber \\
+x_-(s)x_1(s)L_1(s)\Big)-\frac{1}{2}\frac{f_\perp^2}{1+is}\Big(2x_-(s)L_1(s) \nonumber \\
-2L_0(s) +L_R(s)\Big)
\Bigg\}+\zeta_{\Gamma_{\perp}}^{(A)}\big(\{u\},s,q\big) \, .
\end{eqnarray}
The functions $x_-(s)$, $x_1(s)$, $L_0(s)$, $L_1(s)$ and $L_R(s)$ are the same as in (\ref{lr}) - (\ref{l0l1}) with
$v_\perp$ replaced by (\ref{vperpv}). The same is true for $\zeta_{\Gamma_{\perp}}^{(A)}\big(\{u\},s,q\big)$,
which has been defined in (\ref{zetagaape}), and where also (\ref{vperpv}) has been used to replace $v_\perp$ and
$v$.

Performing the limit in the dynamical $\zeta$-function (\ref{zetagammapa}) it reduces to
\begin{equation}\label{zetagammapa00}
\zeta_{\Gamma_{\|}}^{(as)}\big(\{u\},s,q\big)=\zeta_{\Gamma_{\|}}^{(A)}\big(\{u\},s,q\big)
\end{equation}
where $\zeta_{\Gamma_{\|}}^{(A)}\big(\{u\},s,q\big)$ is the model A function (\ref{zetagaapa}) with relation
(\ref{vperpv}) inserted into (\ref{TA}).

Finally the function $X_2$ in (\ref{x2}) simplifies for vanishing
time scale ratios to
\begin{equation}\label{x200}
X_2^{(as)}\big(f_\perp\big)=\frac{f_\perp^2}{2} \, .
\end{equation}
Inserting this expression into (\ref{qfunc}) and (\ref{zetalad}), the dynamical $\zeta$-function (\ref{zetasep})
reads
\begin{eqnarray}\label{zetalambda00}
\zeta_\lambda^{(as)}\big(\{\gamma\},f_\perp\big)=\gamma_\perp^2+\frac{1}{2}\gamma_\|^2
-\frac{f_\perp^2}{2}\left(1+\frac{f_\perp^2}{4}\right) \, .
\end{eqnarray}
The value of $v_\perp$ at the FP depends on how $w_\perp^\prime$ goes to zero in the critical limit
$l\to 0$ compared to $w_\perp^{\prime\prime}$. There are three possible scenarios:

i) $w_\perp^{\prime\prime}$ goes  to zero faster than
$w_\perp^\prime$ so that $s \to 0$. Then $v_\perp$ is turning to the
real value $1$.

ii) $w_\perp^\prime$ and $w_\perp^{\prime\prime}$ behave in the same way so that the ratio
$s=s_0$ is constant. $v_\perp$ is in this case a complex constant
\begin{equation}\label{wppcwp}
v_\perp=\frac{1+is_0}{1-is_0} \, .
\end{equation}

iii) $w_\perp^\prime$ goes  to zero faster than
$w_\perp^{\prime\prime}$ so that $s \to \infty$. Then $v_\perp$ is
turning to the real value $-1$.

The third of the three scenarios above can be excluded from the
discussion because some of the $\zeta$-functions do not stay finite
for $v_\perp=-1$. Finite $\zeta$-functions at the FP and therefore
well defined critical exponents may be obtained only in the first
two scenarios.

The $\zeta$-function for scenario ii) are already given in (\ref{zetagammape00}) - (\ref{zetalambda00}) when
(\ref{wppcwp}) is inserted.

{\bf Case} {\boldmath$s=0:$}

For $v_\perp=1$ ($s=0$) the $\zeta$-functions (\ref{zetagammape00}) and (\ref{zetagammapa00})
simplify to
\begin{eqnarray}\label{zetagammape001}
\zeta_{\Gamma_{\perp}}^{(as0)}\big(\{u\},q,f_\perp\big)=-f_\perp^2\Bigg\{1
-\frac{f_\perp^2}{2}\Big(\frac{27}{2}\ln\frac{4}{3}-3\Big) \Bigg\}   \nonumber \\
+\zeta_{\Gamma_{\perp}}^{(A0)}\big(\{u\},q\big) \, ,
\end{eqnarray}
\begin{equation}\label{zetagammapa001}
\zeta_{\Gamma_{\|}}^{(as0)}\big(\{u\},q\big)=\zeta_{\Gamma_{\|}}^{(A0)}\big(\{u\},q\big)
\, .
\end{equation}
The model A functions (\ref{zetagaape}) and (\ref{zetagaapa}) are now
\begin{eqnarray}\label{zetagammaape001}
\zeta_{\Gamma_{\perp}}^{(A0)}\big(\{u\},q\big)=\frac{u_\perp^2}{9}\left(3\ln\frac{4}{3}-\frac{1}{2}\right)
\nonumber \\
+\frac{u_\times^2}{36}\left(L_\perp^{(\times)}(q)-\frac{1}{2}\right)
\end{eqnarray}
and
\begin{eqnarray}\label{zetagammaapa001}
\zeta_{\Gamma_{\|}}^{(A0)}\big(\{u\},q\big)=\frac{u_\|^2}{4}\left(\ln\frac{4}{3}-\frac{1}{6}\right)
\nonumber \\
+\frac{u_\times^2}{18}\left(L_\|^{(\times)}(q)-\frac{1}{2}\right) \,
,
\end{eqnarray}
where in $L_i^{(\times)}(q)$, introduced in (\ref{lperp}) and (\ref{TA}), the relations
(\ref{vperpv}) with $s=0$ have been inserted.

\subsection{Fixed points in the asymptotic subspace}

Inserting the $\zeta$-functions of the previous subsection into
(\ref{betawperps}) - (\ref{betafperps}) one obtains the FP values in
the asymptotic subspace for $s^\star$ finite, or $s^\star=0$. The
results are presented in Tab.\ref{tableI} for the biconical (${\cal
B}$) and the Heisenberg (${\cal H}$) FP. It turns out that
especially at the biconical FP the values of the ratio $q$ are
extremely small, but definitely not zero. Thus the asymptotic
critical behavior in two loop order changes considerably compared to
one loop (see section \ref{onelooporder}). Weak scaling as discussed
in section \ref{generalbehavior} is valid. The two order parameters
scale with the same dynamic exponent $z_{OP}$ from relation
(\ref{zweakop}), while the secondary density scales with a different
dynamic exponent $z_m$ given in (\ref{zweaksd}). The numerical
values of these two dynamic exponents are also given in
Tab.\ref{tableI} in two loop order. Note that the
values of the dynamical exponents to the accuracy shown are independent wether the FP
value of $s$ is zero or not.
\begin{table}[h,t]
\centering \tabcolsep=2.5mm
\begin{tabular}{cccccc}\hline \hline
       & $f_\perp^\star$ & $q^{\star}$ &$s^{\star}$ &$z_{OP}$ &  $z_m$ \\ \hline
C   \cite{C} & - &  -  & $0$& $2.18$ & $2.18$\\
F \cite{F}&  $0.83$ & - & $0$ & $\sim 1.5$  & $\sim 1.5$ \\
${\cal B}$ & $1.232$ & $1.167\cdot 10^{-86}$ & $0$ & $2.048$   & $1.131$ \\
${\cal H}$ &  $1.211$ &  $3.324\cdot 10^{-8}$ & $0$ & $2.003$ & $1.542$ \\
${\cal B}$ & $1.232$ & $2.51\cdot 10^{-782}$ & $0.705$ & $2.048$     & $1.131$ \\
${\cal H}$ &  $1.211$ &  $3.16\cdot 10^{-66}$ & $0.698$ & $2.003$   & $1.542$ \\
\hline\hline
\end{tabular}
\caption{FP values of the mode coupling $f_\perp$ and the ratios
$q=w_\|/w_\perp^\prime$ and
$s=w_\perp^{\prime\prime}/w_\perp^\prime$  in the subspace $w_\|=0$,
$w_\perp=0$ and finite $v=q/(1+is)$ for different cases of the
biconical ${\mathcal B}$ and Heisenberg ${\mathcal H}$ FP in $d=3$.  For
comparison results for model C and model F FPs are shown at
$n=1$ and $n=2$, correspondingly. \label{tableI}}
\end{table}

The comparison with the dynamical critical exponents in the cases
when the OPs decouple statically and dynamically into model C and
model F shows the changes in the multicritical case where the
exponents are changed but each component reflects the decoupled
values accordingly.

\subsection{Effective exponents in the asymptotic subspace \label{assub}}

The flow of the parameters $q$, $s$ and $f_\perp$ can be found by solving the equations
\begin{eqnarray}
\label{dqdl}
l\frac{d q}{dl}&=&q\left(\zeta_{\Gamma_\|}^{(as)}-\Re[\zeta_{\Gamma_\perp}^{(as)}]
+s\Im[\zeta_{\Gamma_\perp}^{(as)}]\right)
\ , \\
\label{dsdl}
l\frac{d s}{dl}&=&(1+s^2)\Im[\zeta_{\Gamma_\perp}^{(as)}] \ , \\
\label{dfperpdl}
l\frac{d f_\perp}{dl}&=&-\frac{f_\perp}{2}\Big(\varepsilon+\zeta_\lambda^{(as)}-2\zeta_m
+\Re[\zeta_{\Gamma_\perp}^{(as)}]   \nonumber \\
&&-s\Im[\zeta_{\Gamma_\perp}^{(as)}]\Big) .
\end{eqnarray}
\begin{figure}[t]
      \centering{\epsfig{file=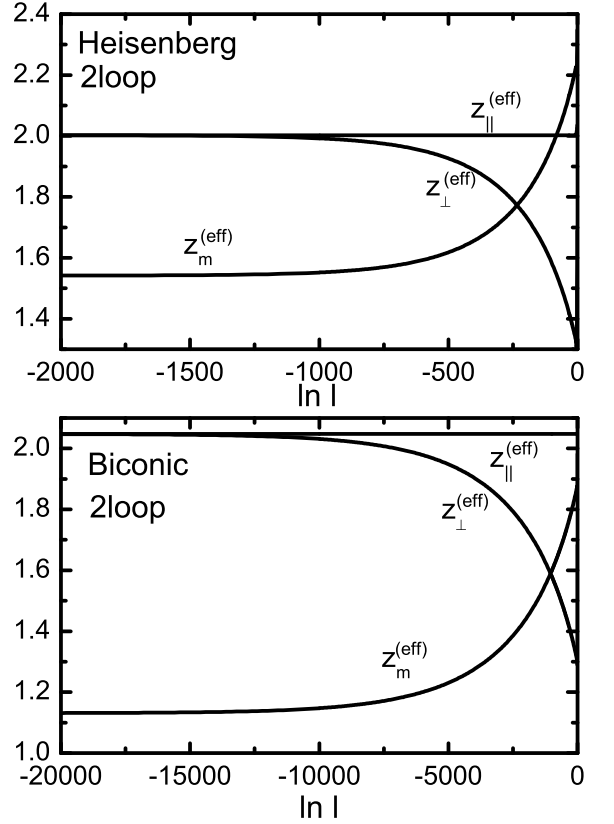,width=9.5cm,angle=0}}
\caption{ \label{zetasub} Effective dynamic exponents in the
subspace $w_\|=w_\perp=0$ with $q$ and $s$ finite in $d=3$. The static values
are taken for the Heisenberg FP and for the biconical FP. The
non-asymptotic region is extended by a factor 10 at the biconical
FP. For the static FP values see Tab. \ref{tab3}, for the dynamic FP values Tab. \ref{tableI}.}
\end{figure}
The $\zeta$-functions in the above flow equations are the reduced expressions (\ref{zetagammape00}),
(\ref{zetagammapa00}) and (\ref{zetalambda00}), which are functions of $q,s,f_\perp$.
We consider the case $s\ne 0$ since the FP $s^\star=0$ is reached only starting with $s=0$.
From the solution of
Eqs.(\ref{dsdl})-(\ref{dfperpdl}) the flow $q(l)$, $s(l)$, $f_\perp(l)$ is obtained, which is used to calculate
asymptotic effective dynamic exponents
\begin{eqnarray}
\label{zperpeffas}
z_\perp^{(as)}(l)&=&2+\Re\big[\zeta_{\Gamma_\perp}^{(as)}\big(q(l),s(l),f_\perp(l)\big)\big] \nonumber \\
&&-s(l)\Im\big[\zeta_{\Gamma_\perp}^{(as)}\big(q(l),s(l),f_\perp(l)\big)\big]
\ , \\
\label{zparaeffas}
z_\|^{(as)}(l)&=&2+\zeta_{\Gamma_\|}^{(as)}\big(q(l),s(l)\big) \ , \\
\label{zmeffas}
z_m^{(as)}(l)&=&2+\zeta_\lambda^{(as)}\big(q(l),s(l),f_\perp(l)\big) \ .
\end{eqnarray}
They can be calculated for different static fixed points, i.e.
biconical or Heisenberg FP, as presented in Fig.\ref{zetasub}. The
values of $u_\perp^\star$, $u_\|^\star$, $u_\times^\star$, as well
as $\gamma_\perp^\star$, $\gamma_\|^\star$, used in the current
calculations can be found in Tab.II. At both fixed
points weak scaling, as discussed in section \ref{generalbehavior},
is fulfilled. The difference to the one loop result is now that the
dynamic exponents $z_\perp$ and $z_\|$ of the OPs are equal in the
asymptotic region, while $z_m$ stays different. Moreover the
transient exponents in two loop order are very small compared to one
loop. There the effective exponents reach their asymptotic values
about $l\sim e^{-15}$ as can be seen from Fig.\ref{z1loop}. In two
loop order the asymptotic region is of magnitudes smaller. From
Fig.\ref{zetasub} one can see that at the Heisenberg FP the flow
parameter has to be of the order $l\sim e^{-1500}$ to obtain the
asymptotic values of the dynamic exponents. At the biconical FP $l$
has to be even of the order $l\sim e^{-15000}$ (note that there is a
factor $10$ between the $x$-scales in Fig.\ref{zetasub}) to reach
asymptotic values. However as will be seen in the next section the
subspace will not be reached by the flow in the complete parameter
space for reasonable values of $l$.

\section{General flow and pseudo-asymptotics}\label{pseudoasymptotics}

Although in general the dynamic flow equations have to be solved
in the full parameter space, the results for the effective
exponents presented in the previous subsection are obtained from the
flow equations which already have been reduced to the subspace
$w_\perp=w_\|=0$. The reason to do this is that the flow and the
$\zeta$-functions in the full parameter space shows some peculiar
behavior.

In the non-asymptotic region the flow is generated by the
system of equations for four parameters which are $w_\perp^\prime$,
$s$, $w_\|$ and $f_\perp$ obtained from Eqs. (\ref{dw1dl}) - (\ref{df1dl}) and (\ref{sq}).
The static parameters are taken at their FP values given in Tab. \ref{tab3}. Due to the presence of the static
asymmetric couplings $\gamma_i$ and the mode coupling $f_\perp$ an
imaginary part of $w_\perp$ is produced even if one starts with a
zero initial value.
\begin{figure}[t]
      \centering{
       \epsfig{file=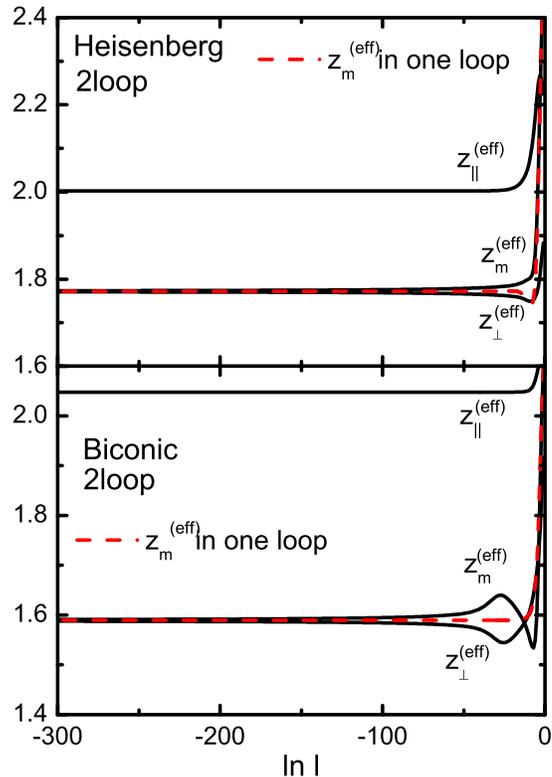,width=9cm,angle=0}}
\caption{ \label{vgl} Effective dynamic exponents in the background using the flow
equations (\ref{dw1dl}),
(\ref{df1dl}) in two loop order in $d=3$ in the complete dynamical parameter space (full lines).
For comparison
the effective dynamic exponent $z_m^{(eff)}$ in one loop order is shown (dashed line).}
\end{figure}
Starting with a typical set of initial values, i.e. $w_\|(l_0)=0.3$,
$w_\perp^\prime(l_0)=0.6$, $s(l_0)=0.5$ and $f_\perp(l_0)=0.4$ at
the flow parameter value $\ln l_0=-1$, the effective exponents in
the complete parameter space have been calculated in $d=3$. The
result is presented in Fig.\ref{vgl} for both static fixed points,
where the solid lines are the results of the two loop calculation
and the dashed line is a result of a complete (flow and effective
exponent) one loop calculation. However the static FP values from
Tab. \ref{tab3} have been used also in the dynamic one loop flow.
There it seems that in two loop order the same results as in the one
loop calculation are obtained. $z_\perp^{eff}$ and $z_m^{eff}$ are
getting close together (solid lines) for flow parameters
$l<e^{-100}$ and seem to coincide even numerically with the
corresponding results in one loop order (dashed line). This is the
type of weak scaling in one loop order, which also can be seen from
Fig.\ref{z1loop} and in qualitative contradiction to the discussion
in the previous sections (see Fig.\ref{zetasub}). But the
examination of the flow of the dynamic parameters reveals a
fundamental difference between the one and two loop calculation. In
one loop order the dynamic parameters $w_\perp^\prime$, $s$, $w_\|$
and $f_\perp$ merge to the FP values when the effective exponents
turns over in their constant asymptotic values, which is presented
in Fig.\ref{flow1loop}. This happens in the region about $l<e^{-15}$
and the dynamic parameters stay constant for all lower flow
parameter values. In two loop order the situation is different.
Although the two loop results for the effective exponents look like
one has reached the asymptotic region (the exponents seem to be
constant), the dynamic parameters in contrast are far from their
asymptotic FP values. This is presented in Fig.\ref{naflow}. The
parameters $w_\perp^\prime$ and $s$ are still increasing and
obviously have not reached a FP value. At the first glance the flow
of $f_\perp$ seems to have reached a FP value (see lowest plot in
Fig.\ref{naflow}). But a closer examination shows that this is not
the case. $f_\perp$ is constantly increasing with a very small slope
as can be seen from the inserted small figure, where both axes has
been enlarged. Actually the set of two loop $\beta$-functions does
not have a zero for finite $w_\perp$ and $w_\|$, and therefore no FP
exists in the parameter region of Fig.\ref{naflow}.
\begin{figure}[t,b]
      \centering{
       \epsfig{file=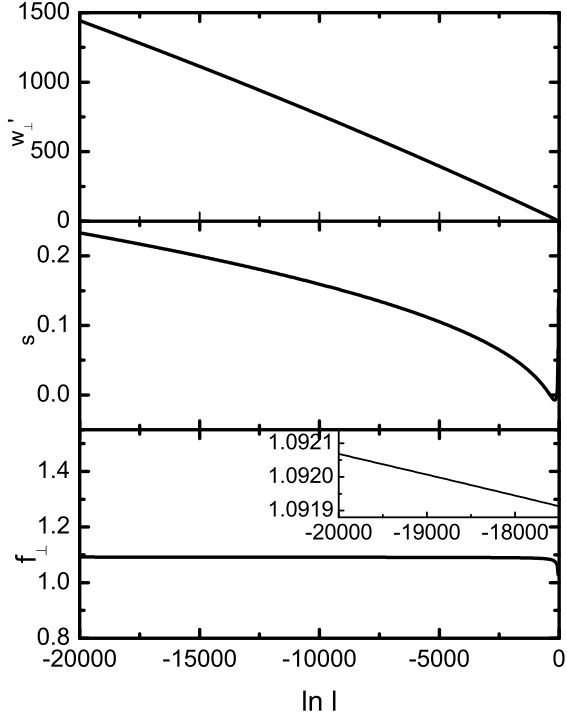,width=8.5cm,angle=0}}
\caption{ \label{naflow} Flow of the parameters $w_\perp^\prime$, $s$ and
$f_\perp$ in the full parameter
space}
\end{figure}
Thus the effective exponents in two loop order in Fig.\ref{vgl} only
show a pseudo-asymptotic behavior completely different from real
asymptotic behavior (there $z_\perp^{eff}$ and $z_\|^{eff}$ have to
be equal) discussed in section \ref{assub} (see Fig. \ref{zetasub}).
Even if one draws the $x$-axis in Fig.\ref{vgl} down to $\ln
l=-20000$, as done for the flow in Fig.\ref{naflow}, the picture
remains to be the same, that is the effective exponents seem to be
constant, with the exception that the background behavior is not
longer visible because the region is too small. Also if one changes
initial conditions of the parameters the qualitative result remains
the same. The different flows merge within a region of $\ln l =-150$
to the same result.

\begin{figure}[t,b]
      \centering{
       \epsfig{file=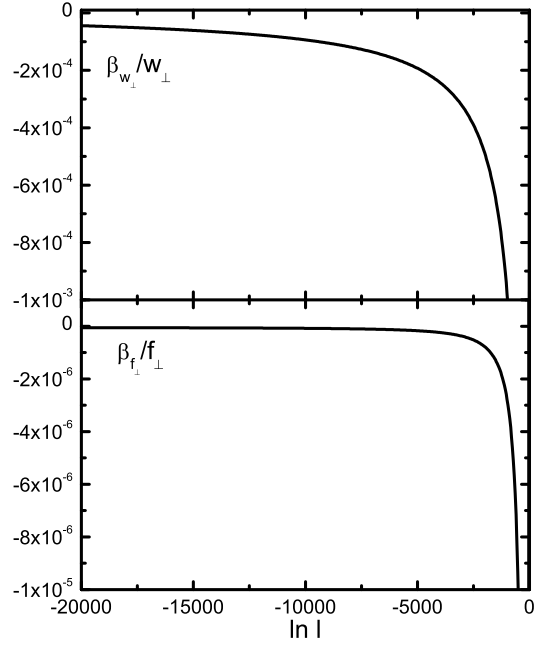,width=8.5cm,angle=0}}
\caption{ \label{flow} Relative slope of the parameters $w_\perp^\prime$
and $f_\perp$ in the full parameter
space}
\end{figure}

In order to get some insight how this pseudo-asymptotic behavior is
possible, in Fig.\ref{flow} the relative slopes
\begin{equation}\label{relslope}
\frac{1}{\alpha_i}\frac{d\alpha_i}{d\ln l}=\frac{\beta_{\alpha_{i}}}{\alpha_i}
\end{equation}
for the parameters $\alpha_i$ chosen to be $w_\perp^\prime$
and $f_\perp$ have been calculated. One can see that the relative
changes in these parameters drop down to very small values. As a
first consequence one has to calculate down to extremely small flow
parameter values $\ln l < -10^8$ where one can expect to leave the
pseudo-asymptotic region. But although the $\beta$-functions cannot
be zero in the considered parameter range they may reach values
which are so small that they cannot longer be separated numerically
from zero. This means that coming from the background it is
impossible to pass the pseudo-asymptotic region numerically into the
real asymptotic region. This is the reason why in the previous
section the flow has been started in an asymptotic subspace.
\begin{figure}[t]
      \centering{
       \epsfig{file=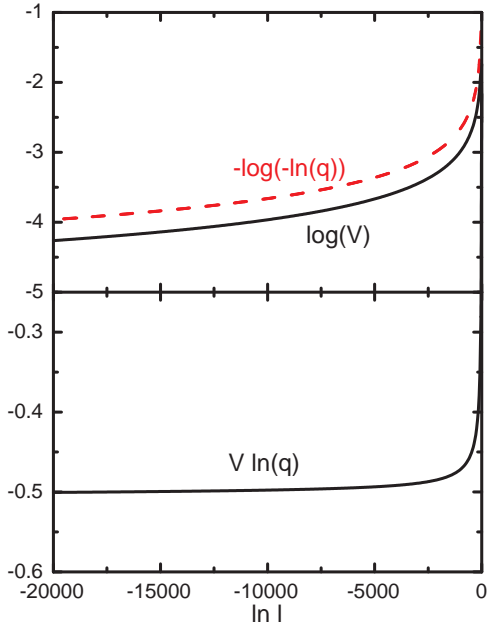,width=8cm,angle=0}}
\caption{ \label{qterms} Contribution of the $V^\prime\ln q$ term in
(\ref{vlnq}) (lower figure). It is of the same magnitude as the
other terms in the $\zeta$-function. $V^\prime$ and $\ln q$ have the
same behavior with opposite exponents. This is demonstrated in the
upper figure, where the decadic logarithm of $V^\prime$ (solid
curve) and the negative decadic logarithm of $-\ln q$ has been drawn
(dashed curves).}
\end{figure}

Due to the presence of logarithmic terms in the timescale ratio $v$
the FP value of $v$ has to be finite. As follows from Tab. \ref{tableI}
via Eq. (\ref{vperpv}) it turns out to be very small leading to
very large (negative) values of $\ln v$. So one expects that the
$\ln v$-terms begin to dominate $\zeta_{\Gamma_\perp}$ in a certain
region of $\ln l$ near the asymptotics. Making the $\ln v$-terms
explicit one may rewrite $\zeta_{\Gamma_\perp}$, given in Eq.
(\ref{zetagammape}), as
\begin{eqnarray}
&&\zeta_{\Gamma_{\perp}}=
-\frac{1}{2}\frac{\gamma_\|D_\perp}{1+w_\perp}\left(\frac{u_\times}{3}
+\frac{1}{2}\frac{\gamma_\|D_\perp}{1+w_\perp}\right)\ln v \nonumber \\
&&-\frac{u_\times^2}{36}\ln v+\mbox{remaining terms} \, .
\end{eqnarray}
Inserting Eq. (\ref{vperpv}), the essential term is the real part
$V^\prime$ of the prefactor of $\ln q$. One obtains
\begin{eqnarray}\label{vlnq}
\zeta_{\Gamma_{\perp}}=V^\prime\ln q+\mbox{remaining terms} \, .
\end{eqnarray}
The prefactor $V^\prime$ is
\begin{equation}\label{vprime}
V^\prime=-\left[\frac{u^2_\times}{36}+\frac{1}{2}\gamma_\|A\left(\frac{u_\times}{3}+\frac{1}{2}
\gamma_\|A\right)-\frac{1}{4}\gamma^2_\|B^2\right]
\end{equation}
with
\begin{equation}\label{vprimea}
A=\frac{w_\perp^\prime[(1+w_\perp^\prime+w_\perp^\prime
s^2)\gamma_\perp-sF]}{(1+w_\perp^\prime)^2 +(w_\perp^\prime s)^2}\,
,
\end{equation}
\begin{equation}\label{vprimeb}
B=\frac{w_\perp^\prime
s\gamma_\perp-(1+w_\perp^\prime)F]}{(1+w_\perp^\prime)^2+(w_\perp^\prime
s)^2} \, .
\end{equation}
In Fig.\ref{qterms} the behavior of the $\ln q$-contributions to
$\zeta_{\Gamma_\perp}$  at the biconical FP is presented. Although
the $\ln q$ term already reaches very large negative values ($q$
is very small), as expected, this is compensated by the prefactor
$V^\prime$ which has very small values in the considered region. In
the upper part of Fig.\ref{qterms} the decadic logarithm of
$V^\prime$ and $-\ln q$ have been plotted. Both curves show a
similar behavior and a small difference. In the lower part
$V^\prime\ln q$ is calculated from (\ref{vprime}) - (\ref{vprimeb}).
As a consequence of the results in the upper part of the figure the
numerical values are about $-0.5$ and one can see that the term is
far away from being the leading one. There are other contributions
to $\zeta_{\Gamma_\perp}$ which have the same magnitude. Thus the
$\zeta$-function is not in the asymptotic region as has been also
indicated by the flow in Fig.\ref{naflow}.

Thus one expects in the experimentally accessible region
non-universal effective dynamical critical behavior. This is
described in the crossover region to the background by the flow
equations together with a suitable matching condition related to the
temperature distance, the wave vector modulus etc. The initial
conditions have to be found by comparison with experiment.

\section{Conclusion}\label{conclusion}
Our two loop calculation for the dynamics at the
multicritical point in anisotropic antiferromagnets in an external
magnetic field leads to a FP where the OPs characterizing the
parallel and perpendicular ordering with respect to the external
field scale in the same way (strong dynamic scaling). This holds independent
wether the Heisenberg FP or the biconical FP in statics is the
stable one. The non-asymptotic analysis of the dynamic flow
equations show that due to cancelation effects the critical behavior
is described - in distances from the critical point accessible to
experiments - by the critical behavior qualitatively found in one
loop order. That means the time scales of the two OP components
 become almost constant in a so called pseudo-asymptotic region and scale differently.

So far we have not included the non-asymptotic flow of the static
parameters which are expected to lead to minor deviations from the
overall picture. Another item would be the study of the decoupled FP
since in the non-asymptotic region the OPs remain statically and
dynamically coupled and the behavior depend on the stability
exponents how fast these effects decay. This in turn depends on the
distance of the system in dimensional space and the space of the OP
components from the stability border line to other FPs than the
decoupled FP (see Fig. 1 in paper I).

The numerical results presented in this series of papers have been calculated for
dimension $d=3$. One might speculate that the peculiar behavior found is specific to the
 dimension of the physical space rather than to the multicritical character of the specific point.
This aspect was out of the scope of this series of papers.
We note that the two critical lines meet at the multicritical point (bicritical or tetracritical)
tangential. This has been taken into account for  the nonsaymptotic behavior by choosing a path approaching the multicritical point without meeting one of the two critical lines \cite{cond}. The nonasymptotic behavior in fact is more complicated since two critical length scales are present in the system. This has to be taken into account when studying the crossover behavior
in approaching one of the critical lines \cite{amit78}.

Only recently a bicritical point has been identified by computer
simulation \cite{selke}. The corresponding FP has been identified as
the Heisenberg FP which corresponds to the type of phase diagram
obtained. It seems to be difficult to look for situations where a
phase diagram containing a tetracritical point is present. Even more
complicated would it be to identify the dynamical characteristic of
this multicritical point, where - coming from the  disordered phase
- two lines belonging to different dynamic universality classes
meet. The dynamical universality class of the case with a $n=1$ OP
(model C) has been studied in \cite{sen,stauffer} with different
results leading to critical exponents larger than expected. The
dynamical universality class of the case with a $n=2$ OP (model F)
case has been studied by computer simulations in
\cite{krech,landau}. The methods of these simulations might be
extended in order to be  used also in the case of the multicritical
point studied in this paper.

{\bf Acknowledgment}\\
This  work was supported by the Fonds zur F\"orderung der
wissenschaftlichen Forschung under Project No. P19583-N20.

Yu.H. acknowledges partial support from the FP7 EU IRSES project  N269139 'Dynamics and Cooperative Phenomena in Complex Physical and Biological Media'.

We thank one of the referees for the exceptionally critical and precise
remarks which have improved the presentation.

\appendix

\section{Calculation of the dynamic vertex functions of the OPs}\label{appop}

In perturbation expansion
up to two loop order the functions $\mathring{\Omega}_{\psi\tilde{\psi}^+}$,
$\mathring{\Gamma}^{(d)}_{\psi\tilde{\psi}^+}$ and $\mathring{\Omega}_{\phi_{\|}\tilde{\phi}_{\|}}$,
which appear in (\ref{gapsipsi}) and (\ref{gaphiphi}), can be written as
\begin{eqnarray}\label{ompsipsi}
\mathring{\Omega}_{\psi\tilde{\psi}^+}(\xi_\perp,\xi_\|,k,\omega)=
1+\mathring{\Omega}^{(1L)}_{\psi\tilde{\psi}^+}(\xi_\perp,k,\omega)  \nonumber \\
+\mathring{\Omega}^{(2L)}_{\psi\tilde{\psi}^+}(\xi_\perp,\xi_\|,k,\omega)
\, ,
\end{eqnarray}
\begin{eqnarray}\label{gdpsipsi}
\mathring{\Gamma}^{(d)}_{\psi\tilde{\psi}^+}(\xi_\perp,\xi_\|,k,\omega)=
2\Big[\mathring{\Gamma}_\perp
+\mathring{G}^{(1L)}_{\psi\tilde{\psi}^+}(\xi_\perp,k,\omega)  \nonumber \\
+\mathring{G}^{(2L)}_{\psi\tilde{\psi}^+}(\xi_\perp,\xi_\|,k,\omega)\Big]
\, ,
\end{eqnarray}
\begin{eqnarray}\label{omphiphi}
\mathring{\Omega}_{\phi_{\|}\tilde{\phi}_{\|}}(\xi_\perp,\xi_\|,k,\omega)=
1+\mathring{\Omega}^{(1L)}_{\phi_{\|}\tilde{\phi}_{\|}}(\xi_\|,k,\omega)  \nonumber \\
+\mathring{\Omega}^{(2L)}_{\phi_{\|}\tilde{\phi}_{\|}}(\xi_\perp,\xi_\|,k,\omega)
\, .
\end{eqnarray}
The superscript $(iL)$ indicates the loop order. Of course all
functions considered depend on all model parameters (couplings and
kinetic coefficients), but only the independent lengths $\xi_\perp$,
$\xi_\|$, $k$ and $\omega$ will be mentioned explicitly in the
following. The one loop contributions are
\begin{equation}\label{ompp1l}
\mathring{\Omega}^{(1L)}_{\psi\tilde{\psi}^+}(\xi_\perp,k,\omega)=
\big(\mathring{\Gamma}_\perp\mathring{\gamma}_\perp-\mbox{i}\mathring{g}\big)
\mathring{\gamma}_\perp I_\perp(\xi_\perp,k,\omega)\, ,
\end{equation}
\begin{equation}\label{gdpp1l}
\mathring{G}^{(1L)}_{\psi\tilde{\psi}^+}(\xi_\perp,k,\omega)=
\big(\mathring{\Gamma}_\perp\mathring{\gamma}_\perp-\mbox{i}\mathring{g}\big)
\mbox{i}\mathring{g}I_\perp(\xi_\|,k,\omega)\, ,
\end{equation}
\begin{equation}\label{omppp1l}
\mathring{\Omega}^{(1L)}_{\phi_{\|}\tilde{\phi}_{\|}}(\xi_\|,k,\omega)=
\mathring{\Gamma}_\|\mathring{\gamma}_\|^2  I_\|(\xi_\|,k,\omega) \,
.
\end{equation}
The one loop integrals $I_\perp$ and $I_\|$ in (\ref{ompp1l})-(\ref{omppp1l})
read
\begin{eqnarray}
\label{ifpe}
I_\perp(\xi_\perp,k,\omega)=\int_{k^\prime}\frac{1}{\big(\xi_\perp^{-2}\!
+\!(k\!+\!k^\prime)^2\big)
(-\mbox{i}\omega+\alpha_\perp^\prime)} \, ,  \\
\label{ifpa}
I_\|(\xi_\|,k,\omega)=\int_{k^\prime}\frac{1}{\big(\xi_\|^{-2}\!+\!(k\!+\!k^\prime)^2\big)
(-\mbox{i}\omega+\alpha_\|^\prime)} \, .
\end{eqnarray}
The dynamic propagators $\alpha_\perp^\prime$ and $\alpha_\|^\prime$ are defined as
\begin{eqnarray}
\label{alphpe}
\alpha_\perp^\prime\equiv\mathring{\Gamma}_\perp\big(\xi_\perp^{-2}\!+\!(k\!+\!k^\prime)^2\big)
+\mathring{\lambda}k^{\prime 2} \, , \\
\label{alphpa}
\alpha_\|^\prime\equiv\mathring{\Gamma}_\|\big(\xi_\|^{-2}\!+\!(k\!+\!k^\prime)^2\big)
+\mathring{\lambda}k^{\prime 2} \, .
\end{eqnarray}
The two loop contributions to the dynamic vertex function of the
orthogonal components (\ref{ompsipsi}) and (\ref{gdpsipsi}) have the
structure
\begin{eqnarray}\label{ompp2l}
\mathring{\Omega}^{(2L)}_{\psi\tilde{\psi}^+}(\xi_\perp,\xi_\|,k,\omega)=
\frac{2}{9}\mathring{\Gamma}_\perp\mathring{u}_\perp^2
 \mathring{W}^{(A_{\perp})}_{\psi\tilde{\psi}^+}(\xi_\perp,k,\omega)\nonumber \\
+\frac{1}{18}\mathring{\Gamma}_\perp\mathring{u}_\times^2
\mathring{W}^{(A_{\times})}_{\psi\tilde{\psi}^+}(\xi_\perp,\xi_\|,k,\omega)\nonumber \\
-\frac{2}{3}\big(2\mathring{\Gamma}_\perp\mathring{\gamma}_\perp
-\mbox{i}\mathring{g}\big)\mathring{u}_\perp
\mathring{F}^{(T3_{\perp})}_{\psi\tilde{\psi}^+}(\xi_\perp,k,\omega) \nonumber \\
-\frac{1}{6}\big(2\mathring{\Gamma}_\perp\mathring{\gamma}_\perp
-\mbox{i}\mathring{g}\big)\mathring{u}_\times
\mathring{F}^{(T3_{\times})}_{\psi\tilde{\psi}^+}(\xi_\perp,\xi_\|,k,\omega) \nonumber \\
+\big(\mathring{\Gamma}_\perp\mathring{\gamma}_\perp-\mbox{i}\mathring{g}\big)
\mathring{\gamma}_\perp
\mathring{F}_{\psi\tilde{\psi}^+}(\xi_\perp,\xi_\|,k,\omega)
\end{eqnarray}
and
\begin{eqnarray}\label{gdpp2l}
\mathring{G}^{(2L)}_{\psi\tilde{\psi}^+}(\xi_\perp,\xi_\|,k,\omega)=
-\frac{2}{3}\mathring{\Gamma}_\perp\mathring{u}_\perp\mbox{i}\mathring{g}
\mathring{F}^{(T3_{\perp})}_{\psi\tilde{\psi}^+}(\xi_\perp,k,\omega) \nonumber \\
-\frac{1}{6}\mathring{\Gamma}_\perp\mathring{u}_\times\mbox{i}\mathring{g}
\mathring{F}^{(T3_{\times})}_{\psi\tilde{\psi}^+}(\xi_\perp,\xi_\|,k,\omega) \nonumber \\
+\big(\mathring{\Gamma}_\perp\mathring{\gamma}_\perp-\mbox{i}\mathring{g}\big)
\mbox{i}\mathring{g}\mathring{F}_{\psi\tilde{\psi}^+}
(\xi_\perp,\xi_\|,k,\omega) \, .
\end{eqnarray}
Note that both two loop functions differ only in terms containing the
static fourth order couplings $\mathring{u}_\perp$ and $\mathring{u}_\times$. The
remaining contributions are the same in both functions apart from a factor
$\mathring{\gamma}_\perp$ and $i\mathring{g}$ respectively. The function
$\mathring{F}_{\psi\tilde{\psi}^+}$ is defined as
\begin{eqnarray}\label{fpp}
\mathring{F}_{\psi\tilde{\psi}^+}(\xi_\perp,\xi_\|,k,\omega)\equiv
\big(\mathring{\Gamma}_\perp\mathring{\gamma}_\perp-\mbox{i}\mathring{g}\big)^2
\mathring{F}^{(T4_{\perp})}_{\psi\tilde{\psi}^+}(\xi_\perp,k,\omega) \nonumber \\
+\mathring{F}^{(T5_\perp)}_{\psi\tilde{\psi}^+}(\xi_\perp,k,\omega)
-\mathring{\gamma}_\perp\mathring{F}^{(T3_\perp)}_{\psi\tilde{\psi}^+}
(\xi_\perp,k,\omega)       \nonumber \\
+\frac{1}{2}\left(\mathring{F}^{(T5_\times)}_{\psi\tilde{\psi}^+}(\xi_\perp,\xi_\|,k,\omega)
-\mathring{\gamma}_\|\mathring{F}^{(T3_\times)}_{\psi\tilde{\psi}^+}
(\xi_\perp,\xi_\|,k,\omega)\right)  \nonumber \\
+\mathring{F}^{(T6_\perp)}_{\psi\tilde{\psi}^+}(\xi_\perp,k,\omega)
-\mathring{\gamma}_\perp\mathring{F}^{(T3_\perp)}_{\psi\tilde{\psi}^+}
(\xi_\perp,k,\omega) \, .
\end{eqnarray}
The first two loop contributions in (\ref{ompp2l}) come from the bicritical
model A. The integrals $\mathring{W}^{(A_{i})}_{\psi\tilde{\psi}^+}$ are defined by
\begin{widetext}
\begin{eqnarray}\label{moda}
\mathring{W}^{(A_{\perp})}_{\psi\tilde{\psi}^+}(\xi_\perp,k,\omega)=
\int_{k^\prime}\int_{k^{\prime\prime}}
\frac{1}{\big(\xi_\perp^{-2}{+}(k{+}k^\prime)^2\big)
(\xi_\perp^{-2}{+}k^{\prime\prime 2})
\big(\xi_\perp^{-2}{+}(k^\prime{+}k^{\prime\prime})^2\big)
(-\mbox{i}\omega{+}A_{\perp\perp^{+}\perp})} \, ,     \\
\mathring{W}^{(A_{\times})}_{\psi\tilde{\psi}^+}(\xi_\perp,\xi_\|,k,\omega)=
\int_{k^\prime}\int_{k^{\prime\prime}}
\frac{1}{\big(\xi_\perp^{-2}{+}(k{+}k^\prime)^2\big)
(\xi_\|^{-2}{+}k^{\prime\prime 2})
\big(\xi_\|^{-2}{+}(k^\prime{+}k^{\prime\prime})^2\big)
(-\mbox{i}\omega{+}A_{\perp\|\|})}
\end{eqnarray}
\end{widetext}
with
\begin{eqnarray}\label{apepepe}
A_{\perp\perp^{+}\perp}\equiv
\mathring{\Gamma}_\perp\big(\xi_\perp^{-2}{+}(k{+}k^\prime)^2\big)
+\mathring{\Gamma}_\perp^+(\xi_\perp^{-2}{+}k^{\prime\prime 2})
\nonumber \\
+\mathring{\Gamma}_\perp\big(\xi_\perp^{-2}{+}(k^\prime{+}k^{\prime\prime})^2\big)
\end{eqnarray}
and
\begin{eqnarray}\label{apepapa}
A_{\perp\|\|}\equiv
\mathring{\Gamma}_\perp\big(\xi_\perp^{-2}{+}(k{+}k^\prime)^2\big)
+\mathring{\Gamma}_\|(\xi_\|^{-2}{+}k^{\prime\prime 2})
\nonumber \\
+\mathring{\Gamma}_\|\big(\xi_\|^{-2}{+}(k^\prime{+}k^{\prime\prime})^2\big)
\, .
\end{eqnarray}
The further two loop contributions in (\ref{ompp2l})-(\ref{fpp}) are marked
with superscripts $(Ti)$, which indicate the different graph topologies. The
explicit expressions are
\begin{widetext}
\begin{eqnarray}\label{ft3}
\mathring{F}^{(T3_{\perp})}_{\psi\tilde{\psi}^+}(\xi_\perp,k,\omega)=
\int_{k^\prime}\int_{k^{\prime\prime}}\frac{1}{\big(\xi_\perp^{-2}{+}(k{+}k^\prime)^2\big)
(-\mbox{i}\omega{+}\alpha_\perp^\prime)(-\mbox{i}\omega{+}A_{\perp\perp^{+}\perp})}
\Bigg(\frac{\mathring{\Gamma}_\perp\mathring{\gamma}_\perp
{-}\mbox{i}\mathring{g}}{\xi_\perp^{-2}{+}k^{\prime\prime 2}}
+\frac{\mathring{\Gamma}_\perp^+\mathring{\gamma}_\perp{+}\mbox{i}\mathring{g}}
{\xi_\perp^{-2}{+}(k^\prime{+}k^{\prime\prime})^2}\Bigg) \, ,
\end{eqnarray}
\begin{eqnarray}\label{ft3x}
\mathring{F}^{(T3_{\times})}_{\psi\tilde{\psi}^+}(\xi_\perp,\xi_\|,k,\omega)=
\int_{k^\prime}\int_{k^{\prime\prime}}\frac{2\mathring{\Gamma}_\|\mathring{\gamma}_\|}
{\big(\xi_\perp^{-2}{+}(k{+}k^\prime)^2\big)
(\xi_\|^{-2}{+}k^{\prime\prime 2})
(-\mbox{i}\omega{+}\alpha_\|^\prime)(-\mbox{i}\omega{+}A_{\perp\|\|})}
\, ,
\end{eqnarray}
\begin{eqnarray}\label{ft4}
\mathring{F}^{(T4_{\perp})}_{\psi\tilde{\psi}^+}(\xi_{\perp},k,\omega)=
\int_{k^\prime}\int_{k^{\prime\prime}}\!\!
\frac{1}{\big(\xi_{\perp}^{-2}\!+\!(k\!+\!k^\prime\!
+\!k^{\prime\prime})^2\big)(-\mbox{i}\omega+\alpha_{\perp}^\prime)^2
(-\mbox{i}\omega+\beta_{\perp})} \, ,
\end{eqnarray}
\begin{eqnarray}\label{ft5}
\mathring{F}^{(T5_{\perp})}_{\psi\tilde{\psi}^+}(\xi_{\perp},k,\omega)=
\int_{k^\prime}\int_{k^{\prime\prime}}
\frac{\mathring{\gamma}_{\perp}\mathring{\lambda}k^{\prime 2}
{-}\mbox{i}\mathring{g}[(k^\prime{+}k^{\prime\prime})^2{-}k^{\prime\prime
2}]} {\big(\xi_{\perp}^{-2}{+}(k{+}k^\prime)^2\big)
(-\mbox{i}\omega{+}\alpha_{\perp}^\prime)^2(-\mbox{i}\omega{+}A_{\perp\perp^{+}\perp})}
\Bigg(\frac{\mathring{\Gamma}_{\perp}\mathring{\gamma}_{\perp}
{-}\mbox{i}\mathring{g}}{\xi_{\perp}^{-2}{+}k^{\prime\prime 2}}
+\frac{\mathring{\Gamma}_{\perp}^+\mathring{\gamma}_{\perp}{+}\mbox{i}\mathring{g}}
{\xi_{\perp}^{-2}{+}(k^\prime{+}k^{\prime\prime})^2}\Bigg) \, ,
\end{eqnarray}
\begin{eqnarray}\label{ft5x}
\mathring{F}^{(T5_{\times})}_{\psi\tilde{\psi}^+}(\xi_{\perp},\xi_\|,k,\omega)=
\int_{k^\prime}\int_{k^{\prime\prime}}
\frac{2\mathring{\Gamma}_\|\mathring{\gamma}_\|^2\mathring{\lambda}k^{\prime
2}}
{\big(\xi_{\perp}^{-2}{+}(k{+}k^\prime)^2\big)(\xi_{\|}^{-2}{+}k^{\prime\prime
2})
(-\mbox{i}\omega{+}\alpha_{\perp}^\prime)^2(-\mbox{i}\omega{+}A_{\perp\|\|})}
\, ,
\end{eqnarray}
\begin{eqnarray}\label{ft6}
\mathring{F}^{(T6_{\perp})}_{\psi\tilde{\psi}^+}(\xi_{\perp},k,\omega)=
\!\!\!\!\!&&\!\!\!\!\!\int_{k^\prime}\int_{k^{\prime\prime}}\!\!
\frac{\mathring{\gamma}_{\perp}\mathring{\lambda}k^{\prime\prime 2}
{+}\mbox{i}\mathring{g}[(k{+}k^\prime{+}k^{\prime\prime})^2{-}(k{+}k^\prime)^2)]}
{\big(\xi_{\perp}^{-2}{+}(k{+}k^\prime)^2\big)
(-\mbox{i}\omega{+}\alpha_{\perp}^\prime)(-\mbox{i}\omega{+}\alpha_{\perp}^{\prime\prime})
(-\mbox{i}\omega{+}S_{\perp\perp^{+}\perp})}
\Bigg(\frac{\mathring{\Gamma}_\perp\mathring{\gamma}_\perp
{-}\mbox{i}\mathring{g}}{\xi_\perp^{-2}{+}(k{+}k^\prime{+}k^{\prime\prime})^2}
+\frac{\mathring{\Gamma}_\perp^+\mathring{\gamma}_\perp{+}\mbox{i}\mathring{g}}
{\xi_\perp^{-2}{+}(k{+}k^{\prime\prime})^2}\Bigg)
\nonumber \\
\!\!&+&\!\!\int_{k^\prime}\int_{k^{\prime\prime}}\!\!
\frac{\mathring{\Gamma}_\perp\mathring{\gamma}_\perp{-}\mbox{i}\mathring{g}}
{\big(\xi_\perp^{-2}{+}(k{+}k^\prime)^2\big)
(-\mbox{i}\omega{+}\alpha_\perp^{\prime\prime})(-\mbox{i}\omega{+}\beta_\perp)}
\Bigg(\frac{\mathring{\Gamma}_\perp\mathring{\gamma}_\perp{-}\mbox{i}\mathring{g}}
{-\mbox{i}\omega{+}\alpha_\perp^\prime}
+\frac{\mathring{\gamma}_\perp}{\xi_\perp^{-2}{+}(k{+}k^\prime{+}k^{\prime\prime})^2}
\nonumber \\
\!\!&&\!\!+\frac{\mathring{\gamma}_\perp\mathring{\lambda}
k^{\prime\prime 2}
{+}\mbox{i}\mathring{g}[(k{+}k^\prime{+}k^{\prime\prime})^2{-}(k{+}k^\prime)^2)]}
{\big(\xi_\perp^{-2}{+}(k{+}k^\prime{+}k^{\prime\prime})^2\big)
(-\mbox{i}\omega{+}\alpha_\perp^\prime)}\Bigg) \, ,
\end{eqnarray}
with the dynamic propagators
\begin{equation}\label{betaij}
\beta_\perp\equiv\mathring{\Gamma}_\perp\big(\xi_\perp^{-2}{+}(k{+}k^\prime{+}k^{\prime\prime})^2\big)
+\mathring{\lambda}\big(k^{\prime 2}{+}k^{\prime\prime 2}\big)\, ,
\end{equation}
\begin{equation}\label{as}
S_{\perp\perp^{+}\perp}\equiv\mathring{\Gamma}_\perp\big(\xi_\perp^{-2}{+}(k{+}k^\prime)^2\big)
+\mathring{\Gamma}_\perp^{+}\big(\xi_\perp^{-2}{+}(k{+}k^\prime{+}k^{\prime\prime})^2\big)
+\mathring{\Gamma}_\perp\big(\xi^{-2}_\perp{+}(k{+}k^{\prime\prime})^2\big)
\end{equation}
which are both invariant under an interchange of $k^\prime$ and $k^{\prime\prime}$.

The two loop contributions to the dynamic vertex function of the
parallel component (\ref{omphiphi}) has the structure
\begin{eqnarray}\label{omphph2l}
\mathring{\Omega}^{(2L)}_{\phi_{\|}\tilde{\phi}_{\|}}(\xi_\perp,\xi_\|,k,\omega)&=&
\frac{1}{6}\mathring{\Gamma}_\|\mathring{u}_\|^2
 \mathring{W}^{(A_{\|})}_{\phi_{\|}\tilde{\phi}_{\|}}(\xi_\|,k,\omega)
+\frac{1}{9}\mathring{\Gamma}_\|\mathring{u}_\times^2
\mathring{W}^{(A_{\times})}_{\phi_{\|}\tilde{\phi}_{\|}}(\xi_\perp,\xi_\|,k,\omega)
 -\mathring{\Gamma}_\|\mathring{u}_\|\mathring{\gamma}_\|
\mathring{F}^{(T3_{\|})}_{\phi_{\|}\tilde{\phi}_{\|}}(\xi_\|,k,\omega) \nonumber \\
&-&\frac{2}{3}\mathring{\Gamma}_\|\mathring{u}_\times\mathring{\gamma}_\|
\mathring{F}^{(T3_{\times})}_{\phi_{\|}\tilde{\phi}_{\|}}(\xi_\perp,\xi_\|,k,\omega)
 +\mathring{\Gamma}_\|\mathring{\gamma}_\|^2
\mathring{F}_{\phi_{\|}\tilde{\phi}_{\|}}(\xi_\perp,\xi_\|,k,\omega)
\, .
\end{eqnarray}
The function $\mathring{F}_{\phi_{\|}\tilde{\phi}_{\|}}$ is defined as
\begin{eqnarray}\label{fphph}
\mathring{F}_{\phi_{\|}\tilde{\phi}_{\|}}(\xi_\perp,\xi_\|,k,\omega)&\equiv&
\mathring{\Gamma}_\|^2\mathring{\gamma}_\|^2
\mathring{F}^{(T4_{\|})}_{\phi_{\|}\tilde{\phi}_{\|}}(\xi_\|,k,\omega)
+\frac{1}{2}\left(\mathring{F}^{(T5_\|)}_{\phi_{\|}\tilde{\phi}_{\|}}(\xi_\|,k,\omega)
-\mathring{\gamma}_\|\mathring{F}^{(T3_\|)}_{\phi_{\|}\tilde{\phi}_{\|}}
(\xi_\|,k,\omega)\right)
+\mathring{F}^{(T5_\times)}_{\phi_{\|}\tilde{\phi}_{\|}}(\xi_\perp,\xi_\|,k,\omega)
\nonumber \\
&-&\mathring{\gamma}_\perp\mathring{F}^{(T3_\times)}_{\phi_{\|}\tilde{\phi}_{\|}}
(\xi_\perp,\xi_\|,k,\omega)
+\mathring{F}^{(T6_\|)}_{\phi_{\|}\tilde{\phi}_{\|}}(\xi_\|,k,\omega)
-\mathring{\gamma}_\|\mathring{F}^{(T3_\|)}_{\phi_{\|}\tilde{\phi}_{\|}}(\xi_\|,k,\omega)
\, .
\end{eqnarray}
The integrals $\mathring{W}^{(A_{i})}_{\phi_{\|}\tilde{\phi}_{\|}}$
in the two loop contributions from the bicritical model A  are
\begin{eqnarray}\label{modapa}
\mathring{W}^{(A_{\|})}_{\phi_{\|}\tilde{\phi}_{\|}}(\xi_\|,k,\omega)=
\int_{k^\prime}\int_{k^{\prime\prime}}
\frac{1}{\big(\xi_\|^{-2}{+}(k{+}k^\prime)^2\big)
(\xi_\|^{-2}{+}k^{\prime\prime 2})
\big(\xi_\|^{-2}{+}(k^\prime{+}k^{\prime\prime})^2\big)
(-\mbox{i}\omega{+}A_{\|\|\|})} \, ,     \\
\mathring{W}^{(A_{\times})}_{\phi_{\|}\tilde{\phi}_{\|}}(\xi_\perp,\xi_\|,k,\omega)=
\int_{k^\prime}\int_{k^{\prime\prime}}
\frac{1}{\big(\xi_\|^{-2}{+}(k{+}k^\prime)^2\big)
(\xi_\perp^{-2}{+}k^{\prime\prime 2})
\big(\xi_\perp^{-2}{+}(k^\prime{+}k^{\prime\prime})^2\big)
(-\mbox{i}\omega{+}A_{\|\perp^{+}\perp})}
\end{eqnarray}
with the propagators
\begin{equation}\label{apapapa}
A_{\|\|\|}\equiv
\mathring{\Gamma}_\|\big(\xi_\|^{-2}{+}(k{+}k^\prime)^2\big)
+\mathring{\Gamma}_\|(\xi_\|^{-2}{+}k^{\prime\prime 2})
+\mathring{\Gamma}_\|\big(\xi_\|^{-2}{+}(k^\prime{+}k^{\prime\prime})^2\big)
\end{equation}
and
\begin{equation}\label{apapepe}
A_{\|\perp^{+}\perp}\equiv
\mathring{\Gamma}_\|\big(\xi_\|^{-2}{+}(k{+}k^\prime)^2\big)
+\mathring{\Gamma}_\perp^+(\xi_\perp^{-2}{+}k^{\prime\prime 2})
+\mathring{\Gamma}_\perp\big(\xi_\perp^{-2}{+}(k^\prime{+}k^{\prime\prime})^2\big)
\, .
\end{equation}
The remaining two loop contributions in (\ref{omphph2l}) and
(\ref{fphph}) are
\begin{eqnarray}\label{ct3}
\mathring{F}^{(T3_{\|})}_{\phi_{\|}\tilde{\phi}_{\|}}(\xi_\|,k,\omega)=
\int_{k^\prime}\int_{k^{\prime\prime}}\frac{2\mathring{\Gamma}_\|\mathring{\gamma}_\|}
{\big(\xi_\|^{-2}{+}(k{+}k^\prime)^2\big)
(\xi_\|^{-2}{+}k^{\prime\prime 2})
(-\mbox{i}\omega{+}\alpha_\|^\prime)(-\mbox{i}\omega{+}A_{\|\|\|})}
\, ,
\end{eqnarray}
\begin{eqnarray}\label{ct3x}
\mathring{F}^{(T3_{\times})}_{\phi_{\|}\tilde{\phi}_{\|}}(\xi_\perp,\xi_\|,k,\omega)=
\int_{k^\prime}\int_{k^{\prime\prime}}\frac{1}{\big(\xi_\|^{-2}{+}(k{+}k^\prime)^2\big)
(-\mbox{i}\omega{+}\alpha_\|^\prime)(-\mbox{i}\omega{+}A_{\|\perp^{+}\perp})}
\Bigg(\frac{\mathring{\Gamma}_\perp\mathring{\gamma}_\perp
{-}\mbox{i}\mathring{g}}{\xi_\perp^{-2}{+}k^{\prime\prime 2}}
+\frac{\mathring{\Gamma}_\perp^+\mathring{\gamma}_\perp{+}\mbox{i}\mathring{g}}
{\xi_\perp^{-2}{+}(k^\prime{+}k^{\prime\prime})^2}\Bigg) \, ,
\end{eqnarray}
\begin{eqnarray}\label{ct4}
\mathring{F}^{(T4_{\|})}_{\phi_{\|}\tilde{\phi}_{\|}}(\xi_|,k,\omega)=
\int_{k^\prime}\int_{k^{\prime\prime}}\!\!
\frac{1}{\big(\xi_{\|}^{-2}\!+\!(k\!+\!k^\prime\!
+\!k^{\prime\prime})^2\big)(-\mbox{i}\omega+\alpha_\|^\prime)^2
(-\mbox{i}\omega+\beta_\|)} \, ,
\end{eqnarray}
\begin{eqnarray}\label{ct5}
\mathring{F}^{(T5_{\|})}_{\phi_{\|}\tilde{\phi}_{\|}}(\xi_|,k,\omega)=
\int_{k^\prime}\int_{k^{\prime\prime}}
\frac{2\mathring{\Gamma}_\|\mathring{\gamma}_\|^2\mathring{\lambda}k^{\prime
2}}
{\big(\xi_{\|}^{-2}{+}(k{+}k^\prime)^2\big)(\xi_{\|}^{-2}{+}k^{\prime\prime
2})
(-\mbox{i}\omega{+}\alpha_{\|}^\prime)^2(-\mbox{i}\omega{+}A_{\|\|\|})}
\, ,
\end{eqnarray}
\begin{eqnarray}\label{ct5x}
\mathring{F}^{(T5_{\times})}_{\phi_{\|}\tilde{\phi}_{\|}}(\xi_\perp,\xi_\|,k,\omega)=
\int_{k^\prime}\int_{k^{\prime\prime}}
\frac{\mathring{\gamma}_{\perp}\mathring{\lambda}k^{\prime 2}
{-}\mbox{i}\mathring{g}[(k^\prime{+}k^{\prime\prime})^2{-}k^{\prime\prime
2}]} {\big(\xi_{\|}^{-2}{+}(k{+}k^\prime)^2\big)
(-\mbox{i}\omega{+}\alpha_\|^\prime)^2(-\mbox{i}\omega{+}A_{\|\perp^{+}\perp})}
\Bigg(\frac{\mathring{\Gamma}_{\perp}\mathring{\gamma}_{\perp}
{-}\mbox{i}\mathring{g}}{\xi_{\perp}^{-2}{+}k^{\prime\prime 2}}
+\frac{\mathring{\Gamma}_{\perp}^+\mathring{\gamma}_{\perp}{+}\mbox{i}\mathring{g}}
{\xi_{\perp}^{-2}{+}(k^\prime{+}k^{\prime\prime})^2}\Bigg) \, ,
\end{eqnarray}
\begin{eqnarray}\label{ct6}
\mathring{F}^{(T6_{\|})}_{\phi_{\|}\tilde{\phi}_{\|}}(\xi_\|,k,\omega)=
\int_{k^\prime}\int_{k^{\prime\prime}}\!\!
\frac{\mathring{\Gamma}_\|\mathring{\gamma}_\|^2\mathring{\lambda}k^{\prime\prime 2}}
{\big(\xi_\|^{-2}{+}(k{+}k^\prime)^2\big)
(-\mbox{i}\omega{+}\alpha_\|^\prime)(-\mbox{i}\omega{+}\alpha_\|^{\prime\prime})
(-\mbox{i}\omega{+}S_{\|\|\|})}
\Bigg(\frac{1}{\xi_\|^{-2}{+}(k{+}k^\prime{+}k^{\prime\prime})^2}
+\frac{1}{\xi_\perp^{-2}{+}(k{+}k^{\prime\prime})^2}\Bigg)
\nonumber \\
+\int_{k^\prime}\int_{k^{\prime\prime}}\!\!\frac{\mathring{\Gamma}_\|\mathring{\gamma}_\|^2}
{\big(\xi_\|^{-2}{+}(k{+}k^\prime)^2\big)
(-\mbox{i}\omega{+}\alpha_\|^{\prime\prime})(-\mbox{i}\omega{+}\beta_\|)}
\Bigg(\frac{\mathring{\Gamma}_\|}{-\mbox{i}\omega{+}\alpha_\|^\prime}
+\frac{1}{\xi_\|^{-2}{+}(k{+}k^\prime{+}k^{\prime\prime})^2}\left(
1+\frac{\mathring{\lambda} k^{\prime\prime 2}}
{-\mbox{i}\omega{+}\alpha_\|^\prime}\right)\Bigg) \, . \nonumber \\
\end{eqnarray}
The additional dynamic propagators are
\begin{equation}\label{aspa}
S_{\|\|\|}\equiv\mathring{\Gamma}_\|\big(\xi_\|^{-2}{+}(k{+}k^\prime)^2\big)
+\mathring{\Gamma}_\|\big(\xi_\|^{-2}{+}(k{+}k^\prime{+}k^{\prime\prime})^2\big)
+\mathring{\Gamma}_\|\big(\xi^{-2}_\|{+}(k{+}k^{\prime\prime})^2\big) \nonumber \\
\end{equation}
and
\begin{equation}\label{betapa}
\beta_\|\equiv\mathring{\Gamma}_\|\big(\xi_\|^{-2}{+}(k{+}k^\prime{+}k^{\prime\prime})^2\big)
+\mathring{\lambda}\big(k^{\prime 2}{+}k^{\prime\prime 2}\big) \, .
\end{equation}
The integrals contained in (\ref{moda}) - (\ref{ft6}) and (\ref{modapa}) - (\ref{ct6}) are of the same type as
already has been presented in \cite{fomo06} (see Eqs.(A19) - (A26) in the appendix therein). The $\varepsilon$-poles of
these integrals can be found in Eqs.(C2) - (C9) of the same reference.

\section{Dynamic Z-factors in two loop order}\label{appz}

Within the minimal subtraction scheme of the renormalization group
calculation one has to collect in two loop order the pole terms of
order $1/\varepsilon^2$ and $1/\varepsilon$ in the functions
$\mathring{\Omega}_{\psi\tilde{\psi}^+}$ and
$\mathring{\Gamma}^{(d)}_{\psi\tilde{\psi}^+}$ in (\ref{gapsipsi}).
The resulting dynamic renormalization factors are
\begin{eqnarray}\label{zpsit}
Z_{\tilde{\psi}^*}^{1/2}=&&\!\!\!\!1-\frac{1}{\varepsilon}
\frac{\gamma_\perp D_\perp}{1+w_\perp}
-\frac{1}{\varepsilon}\left[\frac{u_\perp^2}{18}\left(L_0+x_1L_1-\frac{1}{4}\right)
+\frac{u_\times^2}{72}\left(L_\perp-\frac{1}{4}\right)\right] \nonumber \\
&&\!\!\!\!+\frac{1}{4\varepsilon}\left[\frac{2}{3}
\frac{u_\perp(w_\perp\gamma_\perp
+D_\perp)}{w_\perp(1+w_\perp)}A_\perp
+\frac{\gamma_\perp D_\perp}{w_\perp(1+w_\perp)^2}B_\perp
+\frac{\gamma_\|}{2(1+w_\perp)}\left(\frac{u_\times}{3}(w_\perp\gamma_\perp+D_\perp)
+\frac{w_\perp\gamma_\perp\gamma_\|D_\perp}{1+w_\perp}\right)X_\perp\right] \nonumber \\
&&\!\!\!\!+\frac{1}{2\varepsilon^2}\left[
-\frac{w_\perp\gamma_\perp+D_\perp}{1+w_\perp}\left(\frac{2}{3}u_\perp\gamma_\perp
+\frac{u_\times}{6}\gamma_\|\right) +\frac{\gamma_\perp
D_\perp}{(1+w_\perp)^2}
\left(\frac{D_\perp^2}{1+w_\perp}-w_\perp^2\Big(\gamma_\perp^2+\frac{\gamma_\|^2}{2}\Big)
-\frac{f_\perp^2}{2}\right)\right] \,
\end{eqnarray}
\begin{eqnarray}\label{zgammad}
Z_{\Gamma_\perp}^{(d)}=1\!\!\!&-&\!\!\!\frac{1}{\varepsilon}
\frac{\mbox{i}FD_\perp}{w_\perp(1+w_\perp)}
+\frac{1}{4\varepsilon}\left[\frac{2}{3}
\frac{u_\perp\mbox{i}F}{w_\perp(1+w_\perp)}A_\perp
+\frac{\mbox{i}FD_\perp}{w_\perp^2(1+w_\perp)^2}B_\perp
+\frac{\gamma_\|\mbox{i}F}{2(1+w_\perp)}\left(\frac{u_\times}{3}
+\frac{\gamma_\|D_\perp}{1+w_\perp}\right)X_\perp\right] \nonumber \\
\!\!\!&+&\!\!\!\frac{1}{2\varepsilon^2}\left[
-\frac{\mbox{i}F}{1+w_\perp}\left(\frac{2}{3}u_\perp\gamma_\perp
+\frac{u_\times}{6}\gamma_\|\right) +\frac{\mbox{i}F
D_\perp}{w_\perp(1+w_\perp)^2}
\left(\frac{D_\perp^2}{1+w_\perp}-w_\perp^2\Big(\gamma_\perp^2+\frac{\gamma_\|^2}{2}\Big)
-\frac{f_\perp^2}{2}\right)\right] \, .
\end{eqnarray}
The coupling $D_\perp$ and the functions $A_\perp$, and $B_\perp$
and $X_\perp$ are defined in (\ref{dperp})-(\ref{xperp}). The pole
terms of the function
$\mathring{\Gamma}^{(d)}_{\phi_\|\tilde{\phi}_\|^+}$ are collected
in the renormalization factor
\begin{eqnarray}\label{zgammapa}
Z_{\Gamma_\|}=&&\!\!\!\!1+\frac{1}{\varepsilon}
\frac{w_\|\gamma_\|^2}{1+w_\|}
+\frac{1}{\varepsilon}\left[\frac{u_\|^2}{8}\left(\ln\frac{4}{3}-\frac{1}{6}\right)
+\frac{u_\times^2}{36}\left(vT_A-\frac{1}{2}\right)\right] \nonumber \\
&&\!\!\!\!-\frac{1}{4\varepsilon}\Bigg[\frac{w_\|\gamma_\|^2}{1+w_\|}u_\|
\left(1-3\ln\frac{4}{3}\right)
+\left(\frac{w_\|\gamma_\|^2}{1+w_\|}\right)^2\Bigg(\frac{1}{2}\left(1-9\ln\frac{4}{3}\right)
-\frac{w_\|}{1+w_\|}-\frac{1+2w_\|}{1+w_\|}\ln\frac{(1+w_\|)^2}{1+2w_\|}\Bigg)
\nonumber \\
&&\!\!\!\!+\left(\frac{2}{3}u_\times+\frac{w_\|\gamma_\|}{1+w_\|}\gamma_\perp\right)
\Re\left[\frac{T_1}{w_\perp^\prime}\right]-\frac{\gamma_\|F}{2w_\perp^\prime(1+w_\|)}
\Im\left[\frac{T_2}{w_\perp^\prime}\right]\Bigg] \nonumber \\
&&\!\!\!\!+\frac{1}{2\varepsilon^2}\frac{w_\|\gamma_\|}{1+w_\|}\left[
u_\|\gamma_\|+\frac{2}{3}u_\times\gamma_\perp+\frac{\gamma_\|}{1+w_\|}\left(
w_\|\gamma_\|^2\left(\frac{1}{2}-\frac{w_\|}{1+w_\|}\right)+w_\|\gamma_\perp^2
+\frac{f_\perp^2}{2}\right)+\frac{2w_\|\gamma_\|^3}{1+w_\|}\right]
\, .
\end{eqnarray}
The functions $T_1$, $T_2$ and $T_A$ have been introduced in
(\ref{T1})-(\ref{TA}).

The renormalization factor $Z_\lambda$ is identical to the one of model F \cite{dohm85d} with all
parameters of the perpendicular subsystem. With ${\cal Q}$ defined in (\ref{qfunc}), one gets
\begin{eqnarray}\label{zlamdad}
Z_\lambda^{(d)}=1-\frac{1}{\varepsilon}\ \frac{f_\perp^2}{2}\Bigg\{1
+\frac{{\cal Q}}{2}
-\frac{1}{4\varepsilon}\frac{1}{w_\perp^\prime}\left[
\frac{D_\perp^2}{1+w_\perp}+\frac{D_\perp^{+2}}{1+w_\perp^+}\right]
\Bigg\}   \nonumber \\
\end{eqnarray}
\end{widetext}


\begin{thebibliography}{99}
\bibitem{notefp} We use for the values of the static exponents $\phi/\nu$ its expression
defined by the FP values of the asymmetric couplings via $2\phi/\nu-3=\gamma_\perp^{\star 2}
+\gamma_\|^{\star 2}/2$.
\bibitem{cond} For specialists we note that this condition is represented by Eq. (74) in paper I.
\end{thebibliography}
\end{document}